\documentclass[12pt]{article}

\usepackage[dvipsnames]{xcolor}
\usepackage{graphicx}
\usepackage{amsmath}        
\usepackage{amssymb}        
\usepackage{slashed}
\usepackage{bm}
\usepackage{here}
\usepackage{cite}
\def\mydate{January 29, 2026}
\def\ignore#1{{}}

\def\go{\rightarrow}
\def\dd{\partial}

\def\ep{{\epsilon}}

\def\KK{{\rm KK}}

\def\PMNS{{\rm PMNS}}

\def\CP{{\rm CP}}

\def\onehalf{\hbox{$\frac{1}{2}$}}

\def\la{\langle}
\def\ra{\rangle}

\def\diag{{\rm diag}\,}

\def\cD{{\cal D}}

\def\mymat#1#2{\begin{matrix}#1 \cr \noalign{\kern -2pt} #2\end{matrix}}

\def\mynoalign{\noalign{\kern 4pt}}
\def\mysnoalign{\noalign{\kern 3pt}}
\def\mytinynoalign{\noalign{\kern 2pt}}
\def\ignore#1{{}}

\makeatletter
\@addtoreset{equation}{section}
\makeatother

\begin{document}

\thispagestyle{empty}

%%%%% date and PREPRINT NUMBERS %%%%%%
{\small \noindent \mydate   \hfill }

\vskip 2.5cm

%%%%%%%%%%%%%%%%%%% TITLE %%%%%%%%%%%%%%%%%%
\baselineskip=35pt plus 1pt minus 1pt

\begin{center}
{\bf \Large Neutrino oscillations and PMNS matrix}\\ %[12pt]
{\bf \Large in gauge-Higgs unification}\\ %[12pt]
%{\bf \LARGE Neutrino oscillations and PMNS matrix}\\ %[12pt]
%{\bf \LARGE in gauge-Higgs unification}\\ %[12pt]
\end{center}

%%%%%%%%%%%%%%%% AUTHORS %%%%%%%%%%%%%%%%%%%%%%%

%\vspace{.0cm}
\baselineskip=22pt plus 1pt minus 1pt

\vskip 1.5cm

\begin{center}
\renewcommand{\thefootnote}{\fnsymbol{footnote}}
{\bf  Yutaka Hosotani\footnote[1]{hosotani@rcnp.osaka-u.ac.jp}}

\baselineskip=18pt plus 1pt minus 1pt

\vskip 10pt
{\small \it Research Center for Nuclear Physics, The University of Osaka}\\
{\small \it Ibaraki, Osaka 567-0047, Japan}\\
%{\small \it  Ibaraki, Osaka 567-0047, Japan} \\

\end{center}

\vskip 2.cm
\baselineskip=18pt plus 1pt minus 1pt

\begin{abstract}
In  the $SO(5) \times U(1) \times SU(3)$ gauge-Higgs unification in the Randall-Sundrum  (RS) warped  space
neutrino oscillations and the Pontecorvo-Maki-Nakagawa-Sakata  (PMNS) matrix in the $W$ couplings
originate from Majorana mass terms on the ultraviolet brane in the RS space.  
Tiny neutrino masses are generated by an inverse seesaw mechanism.  
The PMNS matrix in the normal ordering with $\delta_{CP}=\pi$ naturally arises consistently 
with the NuFit-6.0 analysis of neutrino oscillation data.
\end{abstract}

\newpage

\baselineskip=20pt plus 1pt minus 1pt
\parskip=0pt

\section{Introduction} 

The standard model (SM) of  $SU(3)_C \times SU(2)_L \times U(1)_Y$ gauge theory has been 
successful in describing  phenomena at low energies, but has a gauge hierarchy problem 
when embedded in a larger theory such as grand unification.
Gauge-Higgs unification (GHU) models have been proposed to solve the gauge hierarchy problem by
unifying the Higgs boson  and gauge fields.
The SM gauge symmetry $SU(2)_L \times U(1)_Y$ is dynamically broken by an Aharonov-Bohm (AB)  phase, 
$\theta_H$, in the fifth dimension.
The Higgs boson appears as a 4D fluctuation mode of  $\theta_H$.  
A finite mass of the Higgs boson is  generated by quantum effects of the AB phase 
$\theta_H$\cite{Hosotani1983, Davies1988, Hosotani1989, Davies1989,  HetrickHo1989, McLachlan1990, Hatanaka1998, 
Hatanaka1999,  Antoniadis2001, Takenaga2002, Kubo2002,   BurdmanNomura2003,  Csaki2003,  Scrucca2003, 
ACP2005, Cacciapaglia2006, Medina2007, HOOS2008, Serone2010, FHHOS2013, Yoon2018b,
GUTinspired2019a, FCNC2020a, GUTinspired2020b}.

Among various GHU models the $SO(5)\times U(1) \times SU(3)$ GHU in the Randall-Sundrum (RS) 
warped  space,   inspired by $SO(11)$ gauge-Higgs grand unification 
models,\cite{SO11GHGU, Furui2016, HosotaniYamatsu2017, HosotaniYamatsu2018}
has been extensively investigated as a realistic model\cite{GUTinspired2019a, FCNC2020a, GUTinspired2020b}.
The grand unified theory (GUT)-inspired GHU yields
nearly the same phenomenology at low energies as the SM, 
and gives many predictions. % to be confirmed by future experiments.
GHU models in the RS warped space predict large parity violation in the couplings of quarks and leptons
to Kaluza-Klein (KK) excited modes of gauge bosons, whose effects can be  seen in 
forward-backward asymmetry in fermion pair production in electron-positron ($e^- e^+$) 
collisions \cite{Funatsu2017a, Yoon2018a, Funatsu2019a, GUTinspired2020c, Irles2024}.
Deviation from the SM is expected in the processes of $W^- W^+$  and single Higgs production 
as well\cite{Funatsu2023a, Yamatsu2023}.
Further, the GUT-inspired GHU predicts a larger  $W$ boson mass than the SM\cite{Wmass2023}, 
just between the  SM value and the value reported by the CDF Collaboration in 2022\cite{CDF2022}.
The Higgs cubic and quartic self-couplings are predicted to be smaller than in the SM by about 10\%
and 40\%, respectively\cite{GUTinspired2020b, YHbook}.

GHU models on an orbifold such as the RS space exhibit the phenomenon of anomaly flow.
The magnitude of gauge anomalies varies as the AB phase $\theta_H$ changes.  
Further there holds a holographic formula for anomaly coefficients,  expressed in terms of the values of
the wave functions of gauge fields on the ultraviolet (UV) and infrared (IR) branes\cite{AnomalyFlow1, AnomalyFlow2, AnomalyFlow3}.
Extension of GHU models to grand unification has been intensively studied
as well\cite{SO11GHGU, Furui2016, HosotaniYamatsu2017, HosotaniYamatsu2018,LimMaru2007, Kojima2011, Englert2020,
Angelescu2022, Maru2022, Angelescu2023, MaruNago2024, MaruNago2025}.
The  status of GHU has been summarized in Ref.\ \cite{YHbook}.

In a previous paper \cite{CKM2025}  the origin of the Cabibbo-Kobayashi-Maskawa (CKM) matrix
in the $W$ couplings  in the GUT-inspired GHU was clarified. Flavor mixing in the quark sector
arose in the down-type quark multiplets, and the observed CKM matrix was reproduced 
within experimental errors.
Flavor-changing neutral currents (FCNCs) in the $Z$ couplings were naturally suppressed as well.
 
In this paper we investigate the problem of flavor mixing in the lepton sector, namely the origin of 
neutrino oscillations in GHU.   In the GUT-inspired GHU tiny neutrino masses are explained by 
an inverse seesaw mechanism in which Majorana mass terms on the UV brane play an important
role.  The mixing in the neutrino mass eigenstates in the $W$ couplings arises from those Majorana mass terms.
It is known that the mixing angles in the Pontecorvo-Maki-Nakagawa-Sakata  (PMNS) matrix 
are much larger than those in the CKM matrix.  Large mixing angles follow from Majorana mass terms.
% without affecting the $W$ couplings of the observed left-handed neutrinos.

In Section 2 the GUT inspired $SO(5) \times U(1)_X \times SU(3)_C$ GHU model with general
flavor mixing in the lepton sector is described.  Equations of motion and boundary conditions satisfied
by lepton field multiplets are given and general form of their wave functions  are derived.
In Section 3 the equations are solved when brane interactions are diagonal in the generation space.
It is shown that neutrino masses naturally become tiny by an inverse seesaw mechanism.
In Section 4 $W$ couplings of leptons are examined.  It will be found that 
the PMNS matrix\cite{Pontecorvo1958, MNS1962, KM1973} in the normal ordering 
with $\delta_{CP}=\pi$ is realized  consistently 
with the NuFit-6.0 analysis of neutrino oscillation data.\cite{NuFit2024}
A summary is given in Section 5.  
Basis functions for wave functions in the RS space are given in Appendix A.
 
% \vskip 2.cm
 
 \section{GUT-inspired GHU} 
 
The GUT inspired $SO(5) \times U(1)_X \times SU(3)_C (\equiv {\cal G})$ GHU  \cite{GUTinspired2019a}
is defined in the RS warped space with the metric \cite{RS1}
\begin{align}
ds^2= G_{MN} \,  dx^M dx^N =
e^{-2\sigma(y)} \eta_{\mu\nu}dx^\mu dx^\nu+dy^2,
\label{RSmetric1}
\end{align}
where $M,N=0,1,2,3,5$, $\mu,\nu=0,1,2,3$, $y=x^5$, $\eta_{\mu\nu}=\mbox{diag}(-1,+1,+1,+1)$,
$\sigma(y)=\sigma(y+ 2L)=\sigma(-y)$, and $\sigma(y)=ky$ for $0 \le y \le L$.
In terms of the conformal coordinate $z=e^{ky}$ ($0 \le y \le L$, $1\leq z\leq z_L=e^{kL}$)
the metric is expressed as 
\begin{align}
ds^2=  \frac{1}{z^2} \bigg(\eta_{\mu\nu}dx^{\mu} dx^{\nu} + \frac{dz^2}{k^2}\bigg)~ .
\label{RSmetric-2}
\end{align}
The bulk region $0<y<L$ is anti-de Sitter spacetime 
with a cosmological constant $\Lambda=-6k^2$.  It is sandwiched by the
UV brane at $y=0$  and the IR brane at $y=L$.  
The warp factor $z_L$ is large.
The KK mass scale is given by $m_{\rm KK}=\pi k/(z_L-1) \simeq \pi kz_L^{-1}$.
Typical values are $z_L \sim 10^{11}$  and $m_\KK  \sim 13\,$TeV.

Gauge fields 
$A_M^{SO(5)}$,  $A_M^{U(1)_X}$ and $A_M^{SU(3)_C}$ of $SO(5) \times U(1)_X \times SU(3)_C$
with gauge couplings  $g_A$,  $g_B$ and $g_S$
satisfy the orbifold boundary conditions (BCs)
\begin{align}
\begin{pmatrix} A_\mu \cr  A_{y} \end{pmatrix} (x,y_j-y) &=
P_{j} \begin{pmatrix} A_\mu \cr  - A_{y} \end{pmatrix} (x,y_j+y)P_{j}^{-1}
\quad (j=0,1), \cr
\noalign{\kern 5pt}
(y_0, y_1) &= (0, L) , 
\label{BC-gauge1}
\end{align}
where   $P_0=P_1 = P_{\bf 5}^{SO(5)} =\mbox{diag} (I_{4},-I_{1} )$ for $A_M^{SO(5)}$ in the vector 
representation and  $P_0=P_1 = P_{\bf 4}^{SO(5)} =\mbox{diag} (I_{2},-I_{2} )$  for 
$A_M^{SO(5)}$ in the spinorial  representation.    $P_0=P_1= 1$ for $A_M^{U(1)_X}$ and $A_M^{SU(3)_C}$.
The orbifold BCs break $SO(5)$ to $SO(4) \simeq SU(2)_L \times SU(2)_R$.
In the following we write $A_M^{SO(5)} = A_M$ and  $A_M^{U(1)_X} = B_M$ when no confusion arises.

 The 4D Higgs boson doublet $\phi_H(x)$ is the zero mode contained in  $A_z = (kz)^{-1} A_y$;
\begin{align}
A_z^{(j5)} (x, z) &= \frac{1}{\sqrt{k}} \, \phi_j (x) u_H (z) + \cdots , ~~
u_H (z) = \sqrt{ \frac{2}{z_L^2 -1} } \, z ~, \cr
\noalign{\kern 5pt}
\phi_H(x) &= \frac{1}{\sqrt{2}} \begin{pmatrix} \phi_2 + i \phi_1 \cr \phi_4 - i\phi_3 \end{pmatrix} .
\label{4dHiggs}
\end{align}
Without loss of generality we assume $\la \phi_1 \ra , \la \phi_2 \ra , \la \phi_3 \ra  =0$, and  
$\la \phi_4 \ra \not= 0$, 
which is related to the AB phase $\theta_H$ in the fifth dimension by
$\la \phi_4 \ra  = \theta_H f_H$ where
\begin{align}
&f_H  = \frac{2}{g_w} \sqrt{ \frac{k}{L(z_L^2 -1)}} ~~,~ g_w = \frac{g_A}{\sqrt{L}} ~.
\label{fH1}
\end{align}

The matter content relevant in discussing neutrino oscillations and the PMNS matrix  is tabulated 
in Table \ref{Tab:matter}.  
Fields $\Psi_{({\bf 1,4})}^{\alpha}$ ($\alpha=1,2,3$) are in the spinor representation of $SO(5)$, and
contain all leptons as zero modes.  
$\Psi_{({\bf 1,4})}^{\alpha}$ satisfy the orbifold BCs 
\begin{align}
&\Psi_{({\bf 1,4})}^{\alpha} (x, y_j - y) = 
- P_{\bf 4}^{SO(5)} \gamma^5 \Psi_{({\bf 1,4})}^{\alpha} (x, y_j + y) ~.
\label{leptonBC1}
\end{align}
In addition there are $SO(5)$ singlet brane fermion fields $\hat \chi^\alpha (x)$
defined on the UV brane.
$\hat \chi^\alpha$ satisfies the Majorana condition  $(\hat \chi^\alpha)^c = \hat \chi^\alpha$:
\begin{align}
\hat \chi^\alpha  = \begin{pmatrix} \xi_\alpha \cr \eta_\alpha \end{pmatrix} , ~~
(\hat \chi^\alpha)^c = \begin{pmatrix} + (\eta_\alpha)^c \cr - (\xi_\alpha)^c \end{pmatrix} 
=e^{i\delta_C} \begin{pmatrix} + \sigma^2 \eta_\alpha^* \cr - \sigma^2 \xi_\alpha^* \end{pmatrix} .
\label{Majorana1}
\end{align}
The bulk mass parameters of $\Psi_{({\bf 1,4})}^{\alpha}$ are denoted as 
$(c_{\ell_1}, c_{\ell_2}, c_{\ell_3}) = (c_e, c_\mu, c_\tau)$.
The action for  $\Psi_{({\bf 1,4})}^{\alpha}$ in the bulk region is given by
\begin{align}
&I_{\rm lepton}^{\rm bulk} = \int d^5 x \, \sqrt{- G} \, 
\sum_{\alpha=1}^3   \overline{\Psi}_{({\bf 1,4})}^{\alpha} \cD (c_{\ell_\alpha}) \Psi_{({\bf 1,4})}^{\alpha} ~,  \cr
\noalign{\kern 5pt}
&\cD (c) = \Gamma^a {e_a}^M \Big( \dd_M 
+ \frac{1}{8} \omega_{bcM} [\Gamma^b, \Gamma^c] \cr
\noalign{\kern 5pt}
&\hskip 2.5cm
- i g_A   A_M  + \frac{i}{2}  g_B  B_M  \Big) - c \,  \sigma' (y) ~,
\label{SO5leptonaction1}
\end{align}
where $A_M$ is in the spinorial representation.
In terms of   ${\check \Psi}_{\alpha} = z^{-2} \Psi_{({\bf 1,4})}^{\alpha}$ 
the action takes the form
\begin{align}
&I_{\rm lepton}^{\rm bulk} = \int d^4 x  \int_1^{z_L} \frac{dz}{k} ~
\sum_\alpha \overline{\check \Psi}_\alpha
\begin{pmatrix} 
- k \hat D_- (c_{\ell_\alpha}) & \sigma^\mu D_\mu \cr 
\bar \sigma^\mu D_\mu & - k \hat D_+ (c_{\ell_\alpha})
\end{pmatrix}   \check \Psi_\alpha ~ , \cr
\noalign{\kern 5pt}
&\quad 
\hat D_\pm (c) = \pm \,  D_z + \frac{c}{z} ~, \cr
&\quad
D_M = \dd_M - i g_A  A_M  + \frac{i }{2} g_B  B_M~.
\label{SO5leptonaction2}
\end{align}
The action for $\hat \chi^\alpha (x)$ is given by
\begin{align}
I_{\rm brane}^\chi = &
\int d^5x \sqrt{- G} \, \delta(y)  \sum_{\alpha, \beta =1}^3 \frac{1}{2}  \Big\{  
\delta_{\alpha \beta} \overline{\hat\chi}{}^\alpha \gamma^\mu\partial_\mu \hat \chi^\beta
 + M_{\alpha \beta}  \overline{\hat \chi}{}^\alpha \hat \chi^{\beta} \Big\} ~.
\label{branechiaction1}
\end{align}
The  $M_{\alpha \beta}$  term represents Majorana masses.
Hermitian nature of the action, $(I_{\rm brane}^\chi)^\dagger = I_{\rm brane}^\chi$, implies that
$M_{\alpha \beta} = M_{ \beta \alpha}^*$.
It follows from the Majorana condition $(\hat \chi^\alpha)^c = \hat \chi^\alpha$ that
$\overline{\hat \chi} {}^\alpha \hat \chi^\beta = i e^{- i\delta_C} \eta_\alpha^t  \sigma^2 \eta_\beta
+ i e^{i \delta_C} (\eta_\alpha^t  \sigma^2 \eta_\beta)^*$.
For fermion fields $\eta_\alpha^t  \sigma^2 \eta_\beta = \eta_\beta^t  \sigma^2 \eta_\alpha$
so that one can set $M_{\alpha\beta}$ to be symmetric $M_{\alpha\beta} = M_{\beta\alpha}$
and real.

 \begin{table}[tbh]
\renewcommand{\arraystretch}{1.2}
\begin{center}
\caption{Lepton multiplets are introduced in three generations ($\alpha = 1, 2, 3$).
In the second column $\big( SU(3)_C, SO(5) \big)_{U(1)_X}$ content is shown.
In the third column $G_{22}=SU(2)_L\times SU(2)_R$ content is shown.
In the column ``Left handed'' (``Right handed'') the parity $(P_0, P_1)$ of the left (right) handed
components is shown.
Brane Majorana fermion fields $\hat \chi^\alpha$ and 
brane scalar field $\hat \Phi_{({\bf 1}, {\bf 4})}$ defined on the UV brane (at $y=0$) are listed at  bottom  rows.
Note that $\la \hat \Phi_{[{\bf 1}, {\bf 2}]} \ra \not= 0$, while $\la \hat \Phi_{[{\bf 2}, {\bf 1}]} \ra = 0$.
}
\vskip 10pt
\begin{tabular}{cccccc}
%\begin{tabular}{|c|c|c|c|c|c|}
\hline \hline 
Field & ${\cal G}$ & $G_{22}$ &Left handed &Right handed &Name\\
\hline
$\Psi_{({\bf 1,4})}^{\alpha}$ &$({\bf 1,4})_{- \frac{1}{2}}$ &$\, [{\bf 2} , {\bf  1}] \,$
&$(+,+)$ &$(-,-)$ &$\nu_e \,~  \nu_\mu ~\,  \nu_\tau$\\
&&&&& $e ~~~ \mu ~~~ \tau$\\
%\cline{3-6}
&&$[{\bf 1} , {\bf  2}]$ 
&$(-,-)$ &$(+,+)$ &$\nu_e' \,~  \nu_\mu' ~\,  \nu_\tau'$\\
&&&&& $e' ~~ \mu' ~~ \tau'$\\
\hline

$\hat \chi^\alpha$ &$({\bf 1,1})_0$ 
&$[{\bf 1} , {\bf  1}]$
&$\cdots$ &$\cdots$ &$\eta_e ~\, \eta_\mu ~\, \eta_\tau$\\
%\hline %\hline
$\hat \Phi_{({\bf 1}, {\bf 4})}$ &$({\bf 1,4})_{\frac{1}{2}}$ 
&$[{\bf 2} , {\bf  1}]$
&$\cdots$ & $\cdots$ &$\hat \Phi_{[{\bf 2}, {\bf 1}]}$\\
%\cline{3-6}
&&$[{\bf 1} , {\bf 2}]$ & $\cdots$ & $\cdots$ &$\hat \Phi_{[{\bf 1}, {\bf 2}]}$\\
\hline \hline
\end{tabular}
\label{Tab:matter}
\end{center}
\end{table}

There is a brane scalar field $\hat \Phi_{({\bf 1}, {\bf 4})} (x)$ on the UV brane at $y=0$, which spontaneously
breaks $SU(2)_R \times U(1)_X$ to $U(1)_Y$ with $\la \hat \Phi_{({\bf 1}, {\bf 4})} \ra \not= 0$.
It has brane interactions with quark and lepton multiplets, which lead to splitting masses of down-type quarks
from those of up-type quarks, and  induce  an inverse seesaw mechanism for neutrinos as is seen below.
We stress that the model is invariant under five-dimensional $SO(5) \times U(1)_X \times SU(3)_C$ gauge
transformations which preserve the orbifold boundary conditions, even with brane fields and brane 
interactions.\cite{GUTinspired2019a}
One may wonder if the condition can be relaxed  such that only $SO(4) \times U(1)_X \times SU(3)_C$
gauge invariance is required on the branes.   The relaxed condition would allow to introduce additional 
$SO(4)$-multiplet fermions and interactions on the branes, which, however, has a danger to spoil anomaly cancellation.

The brane interactions of $\hat \Phi$ with lepton multiplets are given by 
\begin{align}
&%\hspace{-3em}
 I_{\rm brane}^{\rm int}=  \int d^5x\sqrt{- G} \, \delta(y) \, 
\bigg\{ \sum_{\alpha, \beta} \kappa^\ell_{\alpha\beta} \,
\overline{\hat \chi}{}^{\beta}  \check \Phi_{({\bf 1,4})}^\dagger
 \Psi_{({\bf 1,4})}^{\alpha}  + {\rm H.c.} \bigg\}  ~.
\label{BraneInt1}
\end{align}
Here
\begin{align}
&\hat \Phi _{\bf (1,4)} = \begin{pmatrix} \Phi_{\bf [2,1]} \cr \Phi_{\bf [1,2]}  \end{pmatrix}, ~~~
\check \Phi _{\bf (1,4)} = \begin{pmatrix} i \sigma^2  \Phi_{\bf [2,1]}^* \cr - i \sigma^2 \Phi_{\bf [1,2]}^*  \end{pmatrix}, \cr
\noalign{\kern 5pt}
&\la \hat \Phi _{\bf (1,4)} \ra = \begin{pmatrix} 0\cr 0\cr 0\cr w \end{pmatrix} , ~~~
\la \check \Phi _{\bf (1,4)} \ra = \begin{pmatrix} 0\cr 0\cr - w^* \cr 0 \end{pmatrix} .
\label{BraneScalar1}
\end{align}
Note that both $\hat \Phi _{\bf (1,4)}$ and $\check \Phi _{\bf (1,4)}$ transform as a spinor {\bf 4} in $SO(5)$.
With $\la \tilde \Phi _{\bf (1,4)} \ra \not= 0$ there arise brane mass terms
\begin{align}
&I_{\rm brane ~mass} = \int d^5 x \sqrt{- G} \, \delta(y) 
\bigg\{ \frac{m^B_{\alpha\beta} }{\sqrt{k}} \, \overline{\hat \chi} {}^\beta \check \nu_R^{\prime \alpha} 
+  \frac{m^{B *}_{\alpha\beta} }{\sqrt{k}} \, \overline{\check \nu} {}_R^{\prime \alpha} \hat \chi^\beta \bigg\}
\label{branemass1}
\end{align}
where $m^B_{\alpha\beta} = \kappa^\ell_{\alpha\beta} w \sqrt{k}$.

There are two sources for flavor mixing in the lepton sector.  Majorana masses $M_{\alpha\beta}$ in
Eq.\ (\ref{branechiaction1}) and brane masses $m^B_{\alpha\beta}$ in Eq.\ (\ref{branemass1}).
In this paper we restrict ourselves to the case of diagonal brane masses 
$m^B_{\alpha\beta} = \delta_{\alpha\beta} m_{B_\alpha}$ for which explicit solutions for neutrino
oscillations become available.   In this case the phase of $m_{B_\alpha}$ can be absorbed
by redefining  $\Psi^\alpha_{\bf (1,4)}$.  We take $m_{B_\alpha}$ to be real in the following.

Manipulations are simplified in the twisted gauge \cite{Falkowski2007, HS2007}
defined by  %an $SO(5)$ large gauge transformation
\begin{align}
&\tilde A_M (x,z) = \Omega A_M \Omega^{-1} 
- \frac{i}{g_A} \, \Omega \,\dd_M \Omega^{-1} ~, \cr
\noalign{\kern 5pt}
& \Omega (z)  = \exp \Big\{ i \theta (z) T^{45} \Big\}  ~,~~
\theta (z) = \theta_H \, \frac{z_L^2 - z^2}{z_L^2 - 1} ~, 
\label{twisted1}
\end{align}
where $T^{jk}$'s are $SO(5)$ generators and  
$A_M = 2^{-1/2} \sum_{1 \le j<k \le 5} A_M^{(jk)} T^{jk}$.
In the twisted gauge the background field vanishes ($\tilde \theta_H = 0$).
% so that all fields satisfy free equations in the RS space  in the bulk.  
Boundary conditions at the UV brane are modified,  whereas boundary conditions at the IR brane remain 
the same as in the original gauge.  Quantities in the twisted gauge are denoted by the tilde sign $\tilde{~}$.

Charged leptons are zero modes contained in $\Psi^\alpha_{\bf (1,4)}$.
The mass spectrum $m_{\ell^{(n)}} = k \lambda_{\ell^{(n)}}$  ($\ell= e, \mu, \tau$) is determined by
\begin{align}
&S_L (1;  \lambda_{\ell^{(n)}}, c_\ell)  S_R (1;  \lambda_{\ell^{(n)}}, c_\ell) +\sin^2\onehalf \theta_H =0 ~,
%&S_L (1, \lambda, c_q)  S_R (1, \lambda, c_q) +\sin^2\frac{\theta_H}{2}=0 ~.
\label{chargedLepton-mass1}
\end{align}
where  basis functions $S_{L/R} (z, \lambda, c)$ and $C_{L/R} (z, \lambda, c)$ are given by (\ref{functionA2}).
We follow the notation summarized in Ref.\ \cite{YHbook}.
The lowest modes $e^{(0)}$, $\mu^{(0)}$ and $\tau^{(0)}$ are $e$, $\mu$, and $\tau$ leptons.
With $m_e$, $m_\mu$ and $m_\tau$ given, the corresponding bulk mass parameters $c_e$, $c_\mu$ and $c_\tau$ 
are fixed, and subsequently the mass spectra of their KK towers are determined.
For $\theta_H = 0.1$ and $m_\KK = 13\,$TeV, for instance, $k= 1.59 \times 10^{15}\,$GeV, $z_L = 3.83 \times 10^{11}$ and 
$(|c_e|, |c_\mu|, |c_\tau|) = (1.0068,  0.7930, 0.6754)$.
There is no mixing in charged leptons.
 
 The KK expansion of 5D $e (x,z)$ and $e' (x,z)$ fields in the twisted gauge  is given by
\begin{align}
& \begin{pmatrix} \tilde{\check e} \cr \tilde {\check e}' \end{pmatrix} = 
\sqrt{k} \sum_{n=0}^\infty \bigg\{ e^{(n)}_L (x) \begin{pmatrix} f^{e^{(n)}}_L (z) \cr g^{e^{(n)}}_L (z) \end{pmatrix}
+e^{(n)}_R (x) \begin{pmatrix} f^{e^{(n)}}_R (z) \cr g^{e^{(n)}}_R (z) \end{pmatrix} \bigg\} , \cr
\noalign{\kern 5pt}
&\quad
\begin{pmatrix} f^{e^{(n)}}_L (z) \cr g^{e^{(n)}}_L (z) \end{pmatrix}=  \frac{1}{\sqrt{r_{e^{(n)} L}}}
\begin{pmatrix}\bar c_H C_L (z, \lambda_{e^{(n)}}, c_e) \cr
\noalign{\kern 5pt}
- i\bar s_H  \check S_L (z, \lambda_{e^{(n)}}, c_e) \end{pmatrix}, \cr
\noalign{\kern 5pt}
&\quad
\begin{pmatrix} f^{e^{(n)}}_R (z) \cr g^{e^{(n)}}_R (z) \end{pmatrix}=  \frac{1}{\sqrt{r_{e^{(n)} R}}}
\begin{pmatrix} \bar c_H S_R (z, \lambda_{e^{(n)}}, c_e) \cr
\noalign{\kern 5pt}
- i\bar s_H  \check C_R (z, \lambda_{e^{(n)}}, c_e) \end{pmatrix} ,
\label{wave-up1}
\end{align}
where $\check S_L$ and $\check C_R$ are defined in (\ref{functionA2}) and 
$(\bar c_H, \bar s_H) = (\cos \onehalf \theta_H, \sin \onehalf \theta_H)$. 
The normalization factor for each mode is determined by the condition
\begin{align}
\int_1^{z_L} dz \, \Big\{ |f (z) |^2 + |g (z) |^2 \Big\} = 1 \quad {\rm for~} 
\begin{pmatrix} f (z) \cr g (z) \end{pmatrix} .
\label{normalizationF1}
\end{align}
One can show that $r_{e^{(n)} L} = r_{e^{(n)} R} $.  Similar formulas hold for $\mu$ and $\tau$ towers.

For neutrinos $\nu^\alpha$, $\nu^{\prime \alpha}$ and $\hat \chi^{ \alpha}$ intertwine with each other.
Equations of motion in the original gauge are given by
\begin{align}
&\sigma^\mu \dd_\mu \begin{pmatrix} \check \nu^\alpha_L \cr \check \nu^{\prime \alpha}_L \end{pmatrix}
- k \Big( D_- (c_{\ell_\alpha} )  + \frac{i}{2} \theta' (z) \tau^1 \Big) 
 \begin{pmatrix} \check \nu^\alpha_R \cr \check \nu^{\prime \alpha}_R \end{pmatrix} = 0 ~,  \cr
\noalign{\kern 5pt}
&\bar \sigma^\mu \dd_\mu \begin{pmatrix} \check \nu^\alpha_R \cr \check \nu^{\prime \alpha}_R \end{pmatrix}
- k \Big( D_+ (c_{\ell_\alpha} )  -  \frac{i}{2} \theta' (z) \tau^1 \Big) 
 \begin{pmatrix} \check \nu^\alpha_L \cr \check \nu^{\prime \alpha}_L \end{pmatrix} 
 = \frac{m_{B_\alpha}}{\sqrt{k}} \, \delta (y)  \begin{pmatrix} 0 \cr  \eta_{\alpha} \end{pmatrix} ,  \cr
\noalign{\kern 5pt}
&\Big\{ \sigma^\mu \dd_\mu \eta_\alpha - \frac{m_{B_\alpha}}{\sqrt{k}} \, \check \nu_R^{\prime \alpha}
+ M_{\alpha \beta} \, \eta_\beta^c \Big\} \, \delta (y) = 0 ~.
\label{neutrinoEq1}
\end{align}
We use the notation $(\ell_1, \ell_2, \ell_3) = (e, \mu, \tau)$  throughout the paper.
The brane interaction (\ref{branemass1}) gives rise to a $\delta (y)$ term on the righthand side of the equations above.
$\nu^\alpha_R$ and $\nu^{\prime \alpha}_L$ are parity odd at $y=0$.
Because of the brane interaction $\nu^{\prime \alpha}_L$  becomes discontinuous at $y=0$ as
\begin{align}
&\check \nu^{\prime \alpha}_L \big|_{y=\ep} = - \check \nu^{\prime \alpha}_L \big|_{y=-\ep} 
= - \frac{m_{B_\alpha}}{2 \sqrt{k}} \,  \eta_{\alpha}  ~, \cr
\noalign{\kern 5pt}
&\check \nu^\alpha_R \big|_{y=0}  =  0 ~.
\label{neutrinoBC1}
\end{align}
This is seen by integrating the equations in the $y$ coordinate from $- \ep$ to $\ep$ and taking the limit $\ep \go 0$
with $k D_\pm (c) = e^{-\sigma(y)} [ \pm (\dd/\dd y) + c \sigma '(y) ]$ in the $y$ coordinate.

Each eigenmode with a mass $m=k\lambda$ has $x$-dependence given by
\begin{align}
&\check \nu_L^\alpha, \check \nu_L^{\prime\alpha} , \eta_\alpha \propto f_{\pm L} (x, m) ~, ~~
\check \nu_R^\alpha, \check \nu_R^{\prime\alpha} , \eta_\alpha^c \propto f_{\pm R} (x, m) ~, \cr
\noalign{\kern 5pt}
&\sigma^\mu \dd_\mu  f_{\pm L} (x, m) = m f_{\pm R} (x, m) ~, \cr
\noalign{\kern 5pt}
&\bar \sigma^\mu \dd_\mu  f_{\pm R} (x, m) =m f_{\pm L} (x, m) ~, \cr
\noalign{\kern 5pt}
&f_{\pm L} (x, m)^c = e^{i \delta_C} \sigma^2 f_{\pm L} (x, m)^* = \pm  f_{\pm R} (x, m) ~.
\label{fxMajorana}
\end{align}
There are two types of solutions with $f_{\pm L} (x, m)^c =  \pm  f_{\pm R} (x, m)$.
In the twisted gauge the equations in the bulk region $1 < z < z_L$ satisfied by each eigenmode are
\begin{align}
&D_+ (c_{\ell_\alpha}) \begin{pmatrix} \tilde{\check \nu} {}^\alpha_L \cr \tilde{\check \nu} {}^{\prime\alpha}_L \end{pmatrix}
- \lambda  \begin{pmatrix} \tilde{\check \nu} {}^\alpha_R \cr \tilde{\check \nu} {}^{\prime\alpha}_R \end{pmatrix} =0 ~,  \cr
\noalign{\kern 5pt}
&D_- (c_{\ell_\alpha}) \begin{pmatrix} \tilde{\check \nu} {}^\alpha_R \cr \tilde{\check \nu} {}^{\prime\alpha}_R \end{pmatrix}
- \lambda  \begin{pmatrix} \tilde{\check \nu} {}^\alpha_L \cr \tilde{\check \nu} {}^{\prime\alpha}_L \end{pmatrix} =0 ~.
\label{neutrinoEq2}
\end{align}
Boundary conditions satisfied at $z=z_L$ are the same as in the original gauge.
\begin{align}
&D_+ (c_{\ell_\alpha}) \, \tilde{\check \nu} {}^\alpha_L  = 0 ~, ~~\tilde{\check \nu} {}^{\prime\alpha}_L = 0 ~, \cr
\noalign{\kern 5pt}
& \tilde{\check \nu} {}^\alpha_R = 0 ~, ~~ D_- (c_{\ell_\alpha}) \, \tilde{\check \nu} {}^{\prime\alpha}_R = 0 ~.
\label{neutrinoBC2}
\end{align}
Boundary conditions satisfied at $z=1^+$ are found from (\ref{neutrinoBC1}) and the last equation in (\ref{neutrinoEq1}).
\begin{align}
&\bar c_H  \tilde{\check \nu} {}^\alpha_R + i \bar s_H  \tilde{\check \nu} {}^{\prime\alpha}_R = 0 ~, \cr
\noalign{\kern 5pt}
&i \bar s_H  \tilde{\check \nu} {}^\alpha_L + \bar c_H \tilde{\check \nu} {}^{\prime\alpha}_L 
+  \frac{m_{B_\alpha}}{2 \sqrt{k}} \,  \eta_\alpha = 0 ~, \cr
\noalign{\kern 5pt}
&\frac{m_{B_\alpha}}{\sqrt{k}} \, (i \bar s_H  \tilde{\check \nu} {}^\alpha_R + \bar c_H  \tilde{\check \nu} {}^{\prime\alpha}_R)
\pm k \lambda \,  \eta_\alpha^c - M_{\alpha\beta} \,  \eta_\beta^c = 0 ~.
\label{neutrinoBC3}
\end{align}
Each eigenmode satisfying Eq.\ (\ref{neutrinoEq2}) and boundary conditions (\ref{neutrinoBC2})  is expressed as
\begin{align}
&\begin{pmatrix}  \tilde{\check \nu} {}^\alpha_L  (x,z) \cr \tilde{\check \nu} {}^{\prime\alpha}_L (x,z) \cr 
\sqrt{k} \, \eta_\alpha (x) \end{pmatrix} = 
\begin{pmatrix} ~ a_\alpha C_L (z, \lambda, c_{\ell_\alpha}) \cr i b_\alpha S_L (z, \lambda, c_{\ell_\alpha}) \cr
 i h_\alpha \end{pmatrix} f_{\pm L} (x, k \lambda) ~, \cr
\noalign{\kern 5pt}
&\begin{pmatrix}  \tilde{\check \nu} {}^\alpha_R  (x,z) \cr \tilde{\check \nu} {}^{\prime\alpha}_R (x,z) \cr 
\sqrt{k} \, \eta_\alpha^c (x) \end{pmatrix} = 
\begin{pmatrix} ~ a_\alpha S_R (z, \lambda, c_{\ell_\alpha}) \cr i b_\alpha C_R (z, \lambda, c_{\ell_\alpha}) \cr 
\mp i h_\alpha^* \end{pmatrix} f_{\pm R} (x, k\lambda) ~.
\label{neutrinoMode1}
\end{align}
The eigenvalue $\lambda$ and corresponding coefficients $(a_\alpha, b_\alpha, h_\alpha)$ are
determined by the boundary conditions (\ref{neutrinoBC3}).
Overall normalization is determined by
\begin{align}
& \sum_{\alpha=1}^3 \bigg\{ \int_1^{z_L} dz \Big( |a_\alpha |^2 \big\{ C_L^\alpha (z)^2 + S_R^\alpha (z)^2 \big\} 
+ |b_\alpha |^2 \big\{ S_L^\alpha (z)^2 + C_R^\alpha (z)^2 \big\} \Big) + |h_\alpha |^2 \bigg\} = 1
\label{normalizationN1}
\end{align}
where $C_L^\alpha (z) = C_L (z, \lambda, c_{\ell_\alpha})$ etc.
 
\section{Seesaw mechanism} 

We first note that the coefficient $b_\alpha$ in (\ref{neutrinoMode1}) is related to $a_\alpha$
by the first relation in (\ref{neutrinoBC3}) as
\begin{align}
b_\alpha &= \frac{\bar c_H S^\alpha_R}{\bar s_H C^\alpha_R} \, a_\alpha ~, \cr
\noalign{\kern 5pt}
\begin{pmatrix} S^\alpha_R \cr C^\alpha_R \end{pmatrix} & = 
\begin{pmatrix} S_R \cr C_R \end{pmatrix} (1, \lambda, c_{\ell_\alpha}) ~.
\label{wave1}
\end{align}
Consistent solutions are found with $a_\alpha^* = a_\alpha$ and $h_\alpha^* = h_\alpha$.
It is most convenient to express equations in terms of dimensionless quantities;
\begin{align}
&\tilde m_{B_\alpha} = \frac{m_{B_\alpha}}{k} ~, ~~
\tilde M_{\alpha\beta} = \frac{M_{\alpha\beta}}{k} ~.
\label{dimless1}
\end{align}
The second and third relations in (\ref{neutrinoBC3}) are expressed, after eliminating $b_\alpha$
by (\ref{wave1}), as
\begin{align}
K \begin{pmatrix} \vec a \cr \noalign{\kern 4pt}  \bar s_H \, \vec h \end{pmatrix} &= 
\begin{pmatrix}  L &  \frac{1}{2}  C_B \cr  \noalign{\kern 5pt}  - S_B & \lambda I \mp \tilde M \end{pmatrix}
\begin{pmatrix} \vec a \cr \noalign{\kern 4pt} \bar s_H \, \vec h \end{pmatrix} =0 ~,  \cr
\noalign{\kern 8pt}
\vec a &= \begin{pmatrix} a_1 \cr a_2 \cr a_3 \end{pmatrix} , ~~
\vec h = \begin{pmatrix} h_1 \cr h_2 \cr h_3 \end{pmatrix} , 
\label{Kmatrix1}
\end{align}
where
\begin{align}
L ~&= \begin{pmatrix} L_1 &&\cr & L_2 &\cr && L_3 \end{pmatrix}, ~~~L_\alpha = S^\alpha_L S^\alpha_R+ \bar s_H^2 ~, \cr
\noalign{\kern 5pt}
S_B &=  \begin{pmatrix}  S_B^1 &&\cr &  S_B^2 &\cr &&  S_B^3\end{pmatrix}, 
~~~ S_B^\alpha =  \tilde m_{B_\alpha} S_R^\alpha ~,   \cr
\noalign{\kern 5pt}
C_B &=  \begin{pmatrix}  C_B^1 &&\cr & C_B^2 &\cr && C_B^3\end{pmatrix},
~~~ C_B^\alpha =  \tilde m_{B_\alpha} C_R^\alpha ~, 
\label{Kmatrix2}
\end{align}
and $(\tilde M)_{\alpha\beta} = \tilde M_{\alpha\beta}$.
Note that $L_\alpha = 0$ for $\lambda = \lambda_{\ell_\alpha}$, but 
$L_\alpha \not= 0$ for neutrino states.

The spectrum of neutrino towers is determined by $\det K = 0$.
It is straightforward to see $\det K =  \det L \cdot \det  ( \lambda I \mp \tilde M + \onehalf S_B C_B L^{-1} )$
so that 
\begin{align}
\det  ( \lambda I \mp \tilde M + \onehalf S_B C_B L^{-1} ) =0 ~.
\label{detK1}
\end{align}
The relation (\ref{detK1}) gives an inverse seesaw mechanism for light neutrino states.
To see it explicitly, let us first consider the case of a diagonal Majorana matrix, 
$\tilde M_{\alpha\beta} = \delta_{\alpha\beta} \tilde M_\alpha$.
The relation (\ref{detK1})  becomes 
\begin{align}
&\lambda \mp \tilde M_\alpha + \frac{\tilde m_{B_\alpha}^2 S_R^\alpha C_R^\alpha}{2 L_\alpha} = 0 ~.
\label{detK2}
\end{align}
Neutrino masses are small and $\lambda \ll \lambda_{\ell_\alpha}$.  For $c_{\ell_\alpha} < - \onehalf$
\begin{align}
% &L_\alpha \big|_{\lambda=\lambda_{\ell_\alpha}} \sim
% - \frac{\lambda_{\ell_\alpha}^2 z_L^{2 |c_{\ell_\alpha}| + 1}}{4 c_{\ell_\alpha}^2 -1} + \bar s_H^2 ~, \cr
% \noalign{\kern 5pt}
&L_\alpha \sim (\lambda_{\ell_\alpha}^2 - \lambda^2) \,  \frac{ z_L^{2 |c_{\ell_\alpha}| + 1}}{4 c_{\ell_\alpha}^2 -1} 
\sim \lambda_{\ell_\alpha}^2 \,\frac{ z_L^{2 |c_{\ell_\alpha}| + 1}}{4 c_{\ell_\alpha}^2 -1} ~, \cr
\noalign{\kern 5pt}
&S_R^\alpha \sim \lambda \, \frac{z_L^{1 + |c_{\ell_\alpha}| }}{1 + 2 |c_{\ell_\alpha}|} ~,~~~
 C_R^\alpha \sim  z_L^{ |c_{\ell_\alpha}|}  ~,
\label{numass1}
\end{align}
so that the relation with the upper sign $-$ in (\ref{detK2}), namely with $f_{+ L}(x, k \lambda)$ dependence 
in (\ref{neutrinoMode1}), gives
\begin{align}
&m = k \lambda \sim \frac{ m_{\ell_\alpha}^2 M_\alpha}{(|c_{\ell_\alpha}| - \onehalf) \, m_{B_\alpha}^2} ~.
\label{numass2}
\end{align}
With a moderate Majorana mass, say $M_\alpha \sim 10^6\,$GeV ($10^{10}\,$GeV)  small neutrino 
masses $m_\nu \sim 1\,$meV to 10$\,$meV are obtained with appropriate 
$m_{B_\alpha} = 10^6 \,$GeV $\sim 10^9\,$GeV ($10^8 \,$GeV $\sim 10^{11}\,$GeV).
This result was first obtained in the context of gauge-Higgs grand unification in Ref.\ \cite{HosotaniYamatsu2017}, 
see also Ref.\ \cite{HasegawaLim2018}, 
and was elaborated  in the GUT-inspired GHU in Ref.\ \cite{GUTinspired2019a}.
We note that for $c_{\ell_\alpha} > \onehalf$ extra KK modes with MeV scale masses appear, which contradicts
with the observation. 
We take, for  $\theta_H = 0.1$ and $m_\KK = 13\,$TeV, $(c_e, c_\mu, c_\tau) = (-1.0068, - 0.7930, -0.6754)$.

Equation (\ref{detK2}) determines the spectrum of neutrino KK towers.
For $c_{\ell_\alpha} <0$ the spectrum of the KK modes with $f_{+ L/R}(x, k \lambda)$ dependence 
in (\ref{neutrinoMode1}) is given by $\{ \lambda_{\nu_\alpha^{+ (n)}} , n \ge 0\}$, % $(n=0,1,2, \cdots )$, 
where  $\nu_\alpha^{+ (0)}$ corresponds to light neutrino states described above.
The  spectrum of the KK modes with $f_{- L/R}(x, k \lambda)$ dependence 
in (\ref{neutrinoMode1}) is given by $\{ \lambda_{\nu_\alpha^{- (n)}} , n \ge 1\}$, which contains no light mode.
The $\nu_\alpha^{+ (n)}$ and $\nu_\alpha^{- (n)}$ modes ($n \ge 1$) are nearly degenerate in masses.
The masses of $\nu_\alpha^{+ (1)}$ and $\nu_\alpha^{- (1)}$ modes are $O (m_\KK)$.

The formula (\ref{numass2}) is in the category of the so-called inverse seesaw mechanism.
Unlike in the seesaw mechanism of Minkowski-type (type I seesaw)  \cite{Minkowski1977, Yanagida1979, GellMann1980}, 
at least three states intertwine with each other in the inverse seesaw mechanism\cite{Mohapatra1986, MohapatraValle1986}. 
Indeed,  the result  (\ref{numass2}) is mimicked by
three states $\nu_L^\alpha$, $\nu_R^{\prime \alpha \, c}$ and $\eta_\alpha$ in each generation having 
 mass terms of the form
\begin{align}
&\frac{i}{2} (\nu_L^{\alpha \, c \, \dagger}, \nu_R^{\prime \alpha \, \dagger} , \eta_\alpha^{c\,  \dagger})
\begin{pmatrix} 0 & m_{\ell_\alpha} & 0 \cr m_{\ell_\alpha}  & 0 & m_{B_\alpha} \cr
0 &  m_{B_\alpha} & M_\alpha \end{pmatrix}
\begin{pmatrix} \nu_L^\alpha \cr \nu_R^{\prime \alpha \, c} \cr \eta_\alpha \end{pmatrix} + {\rm H.c.} ~, 
\label{seesaw1}
\end{align}
which leads to a tiny neutrino mass $m_{\ell_\alpha}^2 M_\alpha /  m_{B_\alpha}^2$.
We stress that the formula (\ref{numass2}) for light neutrino modes has been derived by solving Eq.\ (\ref{detK2})
for the entire KK modes, but not by truncating it for three states.

Seesaw mechanisms in RS warped space have been intensively investigated in the literature \cite{Shafi2004, Agashe2016}, 
in which SM singlet neutrino fields (denoted as $N_\alpha$) with independent bulk mass parameters $c_{N_\alpha}$ 
are introduced in the bulk in addition to regular neutrino fields $\nu_\alpha$ with $c_{\ell_\alpha}$.
It has been shown in Ref.\ \cite{Agashe2016} that warped seesaw mechanism is physically inverted with
appropriate choice of $c_{\ell_\alpha}$, $c_{N_\alpha}$, and Majorana masses on the UV brane.
In the GUT-inspired GHU in RS warped space the $\nu^{\prime \alpha}$ fields correspond to the $N_\alpha$
fields in Ref.\ \cite{Agashe2016}.  In the first generation, for instance, the $\nu_e$, $e$, $\nu_e^\prime$, and $e'$ fields are in one 
multiplet $\Psi^e_{\bf (1,4)}$ in the spinor representation of $SO(5)$ so that all $\nu_e, e, \nu_e'$ and  $e'$
fields have the same bulk mass parameter $c_e$.   Further the 4D Higgs field is a part of the fifth dimensional component
of $SO(5)$ gauge fields so that the Higgs couplings of lepton fields are regulated by the gauge principle.  
Neutrino masses are given by Eq.\  (\ref{numass2}) in GHU.

The spectrum with general $M_{\alpha\beta}$ can be found as well.   We focus on light neutrino states
with $\lambda \ll \lambda_{\ell_\alpha}$.  
By making use of  (\ref{numass1}),   the $K$ matrix in Eq.\ (\ref{Kmatrix1}) with the upper sign in the right-bottom corner,
$\lambda I - \tilde M \sim - \tilde M$,  is written as 
\begin{align}
&K = \begin{pmatrix} Z_0 & \cr &I \end{pmatrix}
\begin{pmatrix} I & \onehalf \tilde m_B \cr - 2 \lambda \tilde m_B^{-1} F^{-1} & - \tilde M \end{pmatrix}
\begin{pmatrix} \tilde L & \cr & I \end{pmatrix} , \cr
\noalign{\kern 5pt}
&Z_0 = \begin{pmatrix} z_L^{|c_{e}|} && \cr & z_L^{|c_{\mu}|}  & \cr && z_L^{|c_{\tau}|} \end{pmatrix} , ~~~
\tilde L = Z_0^{-1} L ~,  \cr
\noalign{\kern 5pt}
&\tilde m_B = \begin{pmatrix} \tilde m_{B_1} && \cr & \tilde m_{B_2} & \cr && \tilde m_{B_3} \end{pmatrix}, \cr
\noalign{\kern 5pt}
&F = \begin{pmatrix} F_1 && \cr & F_2 & \cr && F_3 \end{pmatrix} ,  ~~~
F_\alpha = \frac{\lambda_{\ell_\alpha}^2}{(|c_{\ell_\alpha}| - \onehalf) \, \tilde m_{B_\alpha}^2} ~.
\label{Kmatrix3}
\end{align}
Hence Eq.\  (\ref{Kmatrix1}) becomes
\begin{align}
& \tilde L \, \vec a + \onehalf \tilde m_B \, \bar s_H  \vec h = 0 ~,  \cr
\noalign{\kern 5pt}
&2 \lambda \, \tilde m_B^{-1} \, F^{-1} \, \tilde L \, \vec a + \tilde M \, \bar s_H  \vec h = 0 ~. 
\label{neutrinoEq3}
\end{align}
Combining the two equations above, one finds
\begin{align}
&F \, \tilde M \, \vec h = \lambda \, \vec h ~.
\label{neutrinoEq4}
\end{align}
Three neutrino masses $m_{\nu_j} = k \lambda_{\nu_j}$ ($j=1,2,3$) are determined as eigenvalues of $F \, M$.
For a diagonal Majorana mass matrix $M_{\alpha\beta} = \delta_{\alpha\beta} M_\alpha$, 
(\ref{neutrinoEq4}) reduces to the formula (\ref{numass2}).
The formula (\ref{neutrinoEq4})  gives a  gauge-Higgs seesaw mechanism for general Majorana masses.

\section{Neutrino oscillations} 

The neutrino spectrum $\{ \lambda_{\nu_j} \}$ and corresponding eigenvectors are determined by
% With  the neutrino spectrum $\{ \lambda_{\nu_j} \}$ found, corresponding eigenvectors are determined by
\begin{align}
&F \, \tilde M \, \vec w_j = \lambda_{\nu_j}   \vec w_j~.
\label{neutrinoEq5}
\end{align}
For each neutrino mass eigenstate the coefficient vector $\vec h_j$ is proportional to $ \vec w_j$, and 
the coefficient vector $\vec a_j$ is in turn determined by (\ref{neutrinoEq3}).
Once $\{ \vec a_j  \}$ is fixed, $W$ couplings of leptons are evaluated,  thereby the PMNS matrix 
$U_{\PMNS}$ being determined.
% To find a relationship between $M = k \tilde M$ and $U_{\PMNS}$, 
It follows from (\ref{neutrinoEq5}) and $\tilde M^\dagger = \tilde M$ that 
\begin{align}
&(\lambda_{\nu_j}^* - \lambda_{\nu_k} )\,  \vec w_j {}^\dagger F^{-1}  \vec w_k = 0 ~. 
\label{relation1}
\end{align}
It implies that $\lambda_{\nu_j}$ is real and 
\begin{align}
& \vec w_j {}^\dagger F^{-1} \vec w_k =  \delta_{jk} \, n_j ~, ~~~ n_j =  \vec w_j {}^\dagger F^{-1} \vec w_j ~.
\label{relation2}
\end{align}
It follows that $\tilde M$ can be expressed as
\begin{align}
&\tilde M 
= F^{-1} \bigg\{ \sum_{j=1}^3 \frac{\lambda_{\nu_j}}{n_j} \, \vec w_j  \otimes \vec w_j {}^\dagger \bigg\} F^{-1} ~.
\label{MajoranaMass1}
\end{align}
As $\tilde M_{\alpha\beta} = \tilde M_{\beta\alpha}$, $\vec w_j$ can be chosen to be real up to an overall phase.

For the $\nu_j$ mode $\vec h_j \propto \vec w_j$ and $\vec a_j = - \onehalf \bar s_H \tilde L^{-1} \tilde m_B \vec h_j$.
For charged leptons $\lambda_{\ell_\alpha} \sim (4 c_{\ell_\alpha}^2-1)^{1/2} \, z_L^{- |c_{\ell_\alpha}| -0.5} \, \bar s_H$ 
so that 
\begin{align}
&a_{j\alpha} \propto q_\alpha w_{j\alpha} ~, ~~~
q_\alpha = \frac{\tilde m_{B_\alpha}}{\lambda_{\ell_\alpha}} \, \frac{(4 c_{\ell_\alpha}^2 -1)^{1/2}}{z_L}  ~.
\label{awRelation1}
\end{align}
The wave functions of the $\nu_j$ mode in Eq.\ (\ref{neutrinoMode1}) are saturated by the $\tilde{\check \nu}{}^\alpha_L$
components, that is,  contributions of other components to the normalization condition Eq.\ (\ref{normalizationN1})
are numerically negligible ($\ll 1$).
The KK expansion of the neutrino fields can be expressed as
\begin{align}
\begin{pmatrix} \tilde{\check \nu}{}^e_L \cr  \tilde{\check \nu}{}^\mu_L \cr  \tilde{\check \nu}{}^\tau_L \end{pmatrix} %(x,z)
&= \sqrt{k} \sum_{j=1}^3  \nu_{jL} (x) % \psi^{\nu_j}_L (x) 
\begin{pmatrix} p_{je} f^{\nu_j}_{eL} (z) \cr  p_{j\mu} f^{\nu_j}_{\mu L} (z) \cr  p_{j\tau} f^{\nu_j}_{\tau L} (z) \end{pmatrix}
+ \cdots ~, \cr
\noalign{\kern 5pt}
f^{\nu_j}_{\alpha L} (z) &=  \frac{1}{\sqrt{\hat r_{j\alpha} }} \, C_L (z, \lambda_{\nu_j}, c_{\ell_\alpha}) ~, ~~~
\hat r_{j\alpha} = \int_1^{z_L} dz \,  C_L (z, \lambda_{\nu_j}, c_{\ell_\alpha})^2 ~,  \cr
\noalign{\kern 5pt}
p_{j\alpha} &= q_\alpha w_{j\alpha} \sqrt{\hat r_{j\alpha}/r_j} ~, ~~~
r_j = \sum_{\alpha=1}^3 q_\alpha^2 \, | w_{j\alpha} |^2 \, \hat r_{j\alpha} ~.
%r_j = \sum_{\alpha=e,\mu, \tau}  q_\alpha^2 \, |w_{j\alpha}|^2 \, \hat r_{j\alpha} ~.
\label{neutrinoKK1}
\end{align}
Once $\vec w_j$ is given, $\vec p_j$ is determined.  Conversely with $\vec p_j$ specified, $\vec w_j$ is
determined up to a normalization factor. 
As neutrino masses are tiny ($m_{\nu_j} < 100\,$meV),  one finds that $\hat r_{j\alpha} \sim \hat r_{\alpha}$.  
For $\theta_H = 0.1$ and $m_\KK = 13\,$TeV, for instance, 
$(\hat r_1,\hat  r_2, \hat r_3) = (\hat r_e,\hat  r_\mu, \hat r_\tau) = (0.3318,0.3867, 0.4254)$.

Gauge couplings of leptons are contained in the covariant derivative $D_M$ in Eq.\ (\ref{SO5leptonaction1}).
% Let $ \{ T^{a_L}, T^{a_R}, T^{\hat b} ; a_L, a_R =1 \sim 3, b= 1 \sim 4 \}$ 
Let $\{ T^\alpha; \alpha=1 \sim 10 \} = \{ T^{a_L}, T^{a_R}, T^{\hat b} ; a=1 \sim 3, b= 1 \sim 4 \}$ 
be generators of $SO(5)$.
The $W$ boson field, $W_\mu (x)$,  is contained in 
$\sum_{a=1}^2 ( A_\mu^{a_L}  T^{a_L} + A_\mu^{a_R}  T^{a_R} + A_\mu^{\hat a} T^{\hat a} )$.
Its KK expansion in the twisted gauge is given by
\begin{align}
&\frac{1}{\sqrt{2 k}} \begin{pmatrix} \tilde A_\mu^{1_L} + i \tilde A_\mu^{2_L} \cr
\tilde A_\mu^{1_R} + i \tilde A_\mu^{2_R} \cr 
\tilde A_\mu^{\hat 1} + i \tilde A_\mu^{\hat 2} \end{pmatrix} 
=  \sum_{n=0}^\infty W_\mu^{ (n)} (x) \begin{pmatrix}  h^L_{W^{(n)}} (z) \cr \mysnoalign
 h^R_{W^{(n)}} (z) \cr  \mysnoalign  \hat h_{W^{(n)}} (z) \end{pmatrix} + \cdots , \cr
\noalign{\kern 5pt}
& \begin{pmatrix}  h^L_{W^{(n)}} (z) \cr \mysnoalign
 h^R_{W^{(n)}} (z) \cr  \mysnoalign  \hat h_{W^{(n)}} (z) \end{pmatrix} =
 \frac{1}{\sqrt{2 \, r_{W^{(n)}}}}
\begin{pmatrix} (1 + c_H) \, C(z; \lambda_{W^{(n)}}) \cr 
(1 - c_H) \, C(z; \lambda_{W^{(n)}})  \cr  
\sqrt{2} \, s_H \check S(z; \lambda_{W^{(n)}}) \end{pmatrix}  , \cr
\noalign{\kern 5pt}
&\quad c_H = \cos \theta_H ~, ~~ s_H = \sin \theta_H ~, 
\label{WbosonWave1}
\end{align}
where the spectrum $m_{W^{(n)}} = k \lambda_{W^{(n)}}$ is determined by
$2 S C' (1; \lambda_{W^{(n)}}) +s_H^2  \lambda_{W^{(n)}}  =0$.
$C(z; \lambda)$,  $S(z; \lambda)$, and $\check S(z; \lambda)$ are defined in Eq.\  (\ref{functionA1}).
Wave functions are normalized by
\begin{align}
&\int_1^{z_L}  \frac{dz}{z} \Big\{ ( |h^L_{W^{(n)}} |^2 + |h^R_{W^{(n)}} |^2 + | \hat h_{W^{(n)}} |^2 \Big\} = 1 ~.
\label{Wnormalization}
\end{align}
The $W$ boson field is $W_\mu (x) = W_\mu^{(0)}  (x)$.

The $W$ interaction in the twisted gauge is evaluated from
\begin{align}
{\cal L}^W_{\rm int} &= - ig_A \sum_\alpha  \int_1^{z_L} \frac{dz}{\sqrt{k}} \, 
 \overline{\tilde{\check \Psi}}{}_{\bf (1,4)}^\alpha 
\gamma^\mu
\sum_{a=1}^2 ( \tilde A_\mu^{a_L}  T^{a_L} + \tilde A_\mu^{a_R}  T^{a_R} + \tilde  A_\mu^{\hat a} T^{\hat a} )
\tilde{\check \Psi}{}_{\bf (1,4)}^\alpha  
\label{Wfermion1}
\end{align}
by inserting the wave functions of leptons and $W$ boson presented above.
For leptons one has 
\begin{align}
{\cal L}^W_{\rm int} &= - i  \frac{g_w}{\sqrt{2}}  \, W_\mu^\dagger \, 
( \bar e_L, \bar \mu_L, \bar \tau_L) \gamma^\mu  \,  \hat g^{W}_L 
\begin{pmatrix} \nu_{1L}  \cr  \nu_{2L}   \cr  \nu_{3L}  \end{pmatrix}   + {\rm H.c.}
% \begin{pmatrix}  \hat g^{Wud}_{L} & \hat g^{Wus}_{L}  &\hat g^{Wub}_{ L} \cr
% \hat g^{Wcd}_{L} & \hat g^{Wcs}_{L}  &\hat g^{Wcb}_{L} \cr
% \hat g^{Wtd}_{L} & \hat g^{Wts}_{L}  &\hat g^{Wtb}_{L} \end{pmatrix} \begin{pmatrix} d_L \cr s_L \cr b_L \end{pmatrix}
 \cr
\noalign{\kern 5pt}
\hat g^{W}_L &= 
\begin{pmatrix}  \hat g^{We 1}_{L} & \hat g^{We2}_{L}  &\hat g^{We3}_{ L} \cr
\hat g^{W\mu 1}_{L} & \hat g^{W\mu 2}_{L}  &\hat g^{W \mu 3}_{L} \cr
\hat g^{W\tau 1}_{L} & \hat g^{W\tau 2}_{L}  &\hat g^{W\tau 3}_{L} \end{pmatrix} ,
\label{Wcoupling1}
\end{align}
where
\begin{align}
&(\hat g^{W}_L)_{\alpha j} =  G_W[ (h^L_{W^{(0)}}, h^R_{W^{(0)}}, \hat h_{W^{(0)}});
( f^{\ell_\alpha}_{L}, g^{\ell_\alpha}_{L}) ,  
      ( p_{j\alpha} f^{\nu_j}_{\alpha L} ,  p_{j\alpha}'  g^{\nu_j}_{\alpha L} ) ] ~, \cr
\noalign{\kern 5pt}
&G_W [(h^L, h^R, \hat h) ; (f_1, g_1), (f_2, g_2) ]  \cr
\noalign{\kern 2pt}
&\quad
= \sqrt{kL} \int_1^{z_L} dz \, \Big\{ h^{L*}   f_1^*   f_2 +  h^{R*}  g_1^*   g_2
+  \frac{i}{\sqrt{2}} \,  \hat h^*  ( f_1^*  g_2 -  g_1^*  f_2 ) \Big\} ~.
\label{Wcoupling2}
\end{align}
Here $f_L^{\ell_1} (z)   = f_L^{e^{(0)}} (z)$ etc.\ and $ p_{j\alpha}'  g^{\nu_j}_{\alpha L} (z)$ is 
the $\tilde{\check \nu}_L^{\prime \alpha}$ 
component of the $\nu_j$ wave function.  
Righthanded couplings are $O(10^{-23})$, and have been omitted above.
Both  $g^{\ell_\alpha}_{L}(z)$ and  $ p_{j\alpha}'  g^{\nu_j}_{\alpha L} (z)$ are negligibly small so that
to high accuracy
\begin{align}
&(\hat g^{W}_L)_{\alpha j}  \simeq p_{j\alpha} \sqrt{kL} \int_1^{z_L} dz \, 
h^{L \, *}_{W^{(0)}}  f^{\ell_\alpha *}_{L} f^{\nu_j}_{\alpha L} 
\equiv p_{j\alpha} \,  \hat g^{0}_{\alpha j}~.
\label{Wcoupling3}
\end{align}
Furthermore it turns out  that $ \hat g^{0}_{\alpha j} \simeq  \hat g^{0}_{\alpha}$ thanks to 
the smallness of neutrino masses.
For $\theta_H=0.1$ and $m_\KK = 13\,$TeV,  for instance, 
$( \hat g^{0}_{e},  \hat g^{0}_{\mu},  \hat g^{0}_{\tau}) = (0.997647, 0.997644, 0.997642)$.
The PMNS matrix \cite{Pontecorvo1958, MNS1962} is given by
\begin{align}
&U_\PMNS = \frac{1}{ \hat g^{0}_{e}} \, \hat g^{W}_L
\simeq \begin{pmatrix} p_{1e} & p_{2e} & p_{3e} \cr  p_{1\mu} & p_{2\mu} & p_{3\mu} \cr 
p_{1\tau} & p_{2\tau} & p_{3\tau} \end{pmatrix} .
\label{PMNS1}
\end{align}

Neutrino oscillations have been observed.  
The  PMNS matrix is parametrized as
\begin{align}
&U_\PMNS = \begin{pmatrix} 1 & 0 & 0 \cr 0 & c_{23} & s_{23} \cr 0 & - s_{23} & c_{23} \end{pmatrix}
\begin{pmatrix} c_{13} & 0 & s_{13} e^{-i \delta_\CP} \cr 0 & 1 & 0 \cr - s_{13} e^{i \delta_\CP} & 0 & c_{13} \end{pmatrix}
\begin{pmatrix} c_{12} & s_{12} & 0 \cr - s_{12} & c_{12} & 0 \cr 0 & 0 & 1 \end{pmatrix} \cdot  {\cal P} ~ , 
\label{PMNS2}
\end{align}
where $c_{ij} = \cos \theta_{ij}$ and $s_{ij} = \sin \theta_{ij}$ with angles $\theta_{ij} \in [0, \pi/2]$, and 
phase $\delta_\CP \in [0, 2 \pi]$.
${\cal P} = \diag (e^{i\alpha_1}, e^{i\alpha_2}, 1)$ contains two additional Majorana phases, 
playing no role in neutrino oscillations.
Phases of $\vec w_j$ and $\vec p_j$ are absorbed in ${\cal P}$ and an overall phase of $U_\PMNS$.
In the normal ordering (NO) $m_{\nu_1} <  m_{\nu_2} < m_{\nu_3}$, whereas in the inverted ordering (IO) 
$m_{\nu_3} < m_{\nu_1} <  m_{\nu_2} $.
In the current scheme only CP conserving phase $\delta_\CP = 0$ or $\pi$ can be realized.

In the NuFit-6.0 global analysis \cite{NuFit2024} both normal and inverted ordering cases have been analyzed.
In the normal ordering case $\delta_\CP \sim  \pi$ is preferred, whereas in the inverted ordering  case 
$\delta_\CP \sim   3 \pi/2$ is preferred.
We are going to show that the current GHU scenario naturally produces $U_\PMNS$ in the NO with $\delta_\CP =  \pi$.

According to the NuFit-6.0 global analysis, in the NO
\begin{align}
&m_{\nu_1} = m_0 ~, ~~ m_{\nu_2} = \sqrt{m_0^2 + \Delta m_{21}^2} ~, ~~ m_{\nu_3} = \sqrt{m_0^2 + \Delta m_{31}^2} ~, \cr
\noalign{\kern 5pt}
& \Delta m_{21}^2 \sim 7.49 \times 10^{-5} \,{\rm eV}^2 ~, ~~  \Delta m_{31}^2 \sim 2.53 \times 10^{-3} \,{\rm eV}^2 ~,
\label{neutrinoMassExp1}
\end{align}
and
\begin{align}
&\sin^2 \theta_{12} \sim 0.307 ~, ~~ \sin^2 \theta_{23} \sim 0.561 ~, ~~ \sin^2 \theta_{13} \sim 0.022 ~, ~~
\delta_\CP \sim \pi ~,  \cr
\noalign{\kern 5pt}
&U_\PMNS ^{\rm exp} \sim \begin{pmatrix} 0.823 & 0.548 & -0.148 \cr
-0.275 & 0.613 & 0.741 \cr 0.497 & -0.569 & 0.655 \end{pmatrix} \equiv 
\begin{pmatrix} && \cr \vec p_1^{~ \rm exp} &  \vec p_2^{~ \rm exp} &  \vec p_3^{~ \rm exp} \cr && \end{pmatrix} .
\label{PMNSexp}
\end{align}
As a typical value we take $m_0 = 1\,$meV.  Then
\begin{align}
&m_{\nu_1} = 1\,{\rm meV}, ~~ m_{\nu_2} = 8.71\,{\rm meV}, ~~m_{\nu_3} = 50.3\,{\rm meV}.
\label{neutrinoMassExp}
\end{align}

Now we present Majorana mass matrices which reproduce (\ref{PMNSexp}) and (\ref{neutrinoMassExp}).
We consider the case with $\theta_H = 0.1$ and $m_\KK = 13\,$TeV, for which 
$k= 1.59 \times 10^{15}\,$GeV,  $z_L = 3.83 \times 10^{11}$ and 
$(c_e, c_\mu, c_\tau) = (-1.0068, - 0.7930, - 0.6754)$.
As seen from (\ref{neutrinoEq3}) and (\ref{MajoranaMass1}), there remains freedom in choosing the values
of $F_\alpha$ in (\ref{Kmatrix3}), or equivalently  brane masses $m_{B_\alpha}$ in Eq.\ (\ref{branemass1}).

\vskip 10pt 
\noindent
Case I:  $F_e = F_\mu = F_\tau$

In this case $F = F_e \,  I$ and the formula (\ref{MajoranaMass1}) reduces to 
\begin{align}
&\tilde M 
= F_e^{-1}  \sum_{j=1}^3 \frac{\lambda_{\nu_j}}{\vec w_j^\dagger \vec w_j} 
\, \vec w_j  \otimes \vec w_j {}^\dagger  ~.
\label{MajoranaMass2}
\end{align}
Further it is found  that $\vec w_j \propto \vec p_j$ to high accuracy.
It is seen that the set of 
\begin{align}
&F_e = F_\mu = F_\tau = 1. \times 10^{-18} ~, \cr
\noalign{\kern 5pt}
&(m_{B_e} , m_{B_\mu}, m_{B_\tau} ) = (6.83 \times 10^5, 1.90 \times 10^8, 4.17 \times 10^9)\, {\rm GeV} ~, \cr
\noalign{\kern 5pt}
&\vec w_j = \vec p_j^{~ \rm exp} ~, \cr
\noalign{\kern 5pt}
&M = k \tilde M = \begin{pmatrix} 4.40 &- 2.83 &-7.20 \cr - 2.83 & 31.0 & 21.2 \cr - 7.20 & 21.2 &24.7 \end{pmatrix}
\times 10^6 \, {\rm GeV} , 
\label{MajoranaMassSet1}
\end{align}
yields  (\ref{PMNSexp}) and (\ref{neutrinoMassExp}).

\vskip 10pt 
\noindent
Case II:  $F_e \not= F_\mu \not= F_\tau$

In this case one needs to take more general $\vec w_j$.  As $\vec w_j$ and $\vec p_j$ are related to each other
by Eq.\ (\ref{neutrinoKK1}), we take
\begin{align}
&w_{je} = p_{je}^{\, \rm exp} ~, \cr
\noalign{\kern 5pt}
&w_{j \mu} = \bigg( \frac{|c_e| + \onehalf}{ |c_\mu| + \onehalf}  \,  \frac{F_\mu}{F_e} \,
\frac{\hat r_{j e}}{\hat r_{j \mu}} \bigg)^{1/2} \, p_{j \mu}^{\, \rm exp} 
\simeq    \sqrt{ \frac{F_\mu}{F_e} } \, p_{j \mu}^{\, \rm exp}  , \cr
\noalign{\kern 5pt}
&w_{j \tau} = \bigg( \frac{|c_e| + \onehalf}{ |c_\tau | + \onehalf}  \,  \frac{F_\tau}{F_e} \,
\frac{\hat r_{j e}}{\hat r_{j \tau}} \bigg)^{1/2} \, p_{j \tau}^{\, \rm exp}
\simeq    \sqrt{ \frac{F_\tau}{F_e} } \, p_{j \tau}^{\, \rm exp}  ~.
\label{wvector1}
\end{align}
Here we have made use of a fact that 
$\big\{ (|c_{e}| + \onehalf)  \hat r_{je} \big\} / \big\{ (|c_{\mu}| + \onehalf)  \hat r_{j\mu} \big\} =1 +  O(10^{-9}) $,  etc. 
For instance, one finds that the set of
\begin{align}
&(F_e,  F_\mu,  F_\tau )= (1, \, 3, \, 10)  \times 10^{-18} ~, \cr
\noalign{\kern 5pt}
&(m_{B_e} , m_{B_\mu}, m_{B_\tau} ) = (6.83 \times 10^5, 1.10 \times 10^8, 1.32 \times 10^9)\, {\rm GeV} ~, \cr
\noalign{\kern 5pt}
&\vec w_1 = \begin{pmatrix} 0.823 \cr- 0.476 \cr 1.57 \end{pmatrix}, ~
\vec w_2 = \begin{pmatrix} 0.548 \cr 1.06 \cr - 1.80 \end{pmatrix}, ~
\vec w_3 = \begin{pmatrix} -0.148 \cr 1.28 \cr 2.07 \end{pmatrix}, ~\cr
\noalign{\kern 5pt}
&M = k \tilde M = \begin{pmatrix} 4.40 &- 1.63 &-2.28 \cr - 1.63 & 10.3 & 3.89 \cr - 2.28 & 3.89 &2.47 \end{pmatrix}
\times 10^6 \, {\rm GeV} , 
\label{MajoranaMassSet2}
\end{align}
reproduces  (\ref{PMNSexp}) and (\ref{neutrinoMassExp}).
Similarly  the set of
\begin{align}
&(F_e,  F_\mu,  F_\tau )= (1, \, 10, \, 100)  \times 10^{-22} ~, \cr
\noalign{\kern 5pt}
&(m_{B_e} , m_{B_\mu}, m_{B_\tau} ) = (6.83 \times 10^7, 6.00 \times 10^9, 4.17 \times 10^{10})\, {\rm GeV} ~, \cr
\noalign{\kern 5pt}
&\vec w_1 = \begin{pmatrix} 0.823 \cr - 0.868 \cr 4.97 \end{pmatrix}, ~
\vec w_2 = \begin{pmatrix} 0.548 \cr 1.94 \cr - 5.69\end{pmatrix}, ~
\vec w_3 = \begin{pmatrix} -0.148 \cr 2.34 \cr 6.55 \end{pmatrix}, ~\cr
\noalign{\kern 5pt}
&M = k \tilde M = \begin{pmatrix} 44.0 &- 8.94 &-7.20 \cr - 8.94 & 3.10 & 6.72 \cr - 7.20 & 6.72 &2.47 \end{pmatrix}
\times 10^9 \, {\rm GeV} , 
\label{MajoranaMassSet3}
\end{align}
reproduces  (\ref{PMNSexp}) and (\ref{neutrinoMassExp}).

It is appropriate to comment on the scales of $m_{B_\alpha}$ and $M$.
The brane mass $m_{B_\alpha}$ arises from  $\la  \Phi _{\bf (1,4)} \ra \not= 0$ ($w \not= 0$)   in the brane 
interaction (\ref{BraneInt1}).  With $w \not= 0$ $SO(4) \times U(1)_X \simeq SU(2)_L \times SU(2)_R \times U(1)_X$
gauge symmetry is spontaneously broken to $SU(2)_L \times  U(1)_Y$.  This breaking scale $w$ is supposed to
be much larger than $m_\KK$.  The constraint for $m_{B_\alpha}$ is $m_\KK \ll m_{B_\alpha} < k$.
Further one can check that the magnitude of Majorana masses   in  (\ref{branechiaction1})
must be larger than $m_\KK$ to have consistent $W$ couplings of leptons.\cite{GUTinspired2019a, YHbook}
The examples presented above, (\ref{MajoranaMassSet1}),  (\ref{MajoranaMassSet2}) and (\ref{MajoranaMassSet3}), 
satisfy these conditions.
There remains freedom in GHU in the choice of the scales of $m_{B_\alpha}$ and $M$.  
If one assumes that $O(m_{B_\alpha}) = O(M)$, one obtains the same scale of Majorana masses as in the ordinary
Minkowski-type (type I) seesaw mechanism as seen from (\ref{numass2}).

\section{Summary} 

In the GUT-inspired gauge-Higgs unification the  mixing of neutrino mass eigenstates  in the $W$ couplings arises
from the Majorana mass matrix $M$.  Tiny neutrino masses are explained by the inverse seesaw mechanism.
Three neutrino masses $\{ m_{\nu_1}, m_{\nu_2}, m_{\nu_3} \}$ are determined by
\begin{align}
&F \, M \, \vec w_j = m_{\nu_j} \, \vec w_j ~, \cr
\noalign{\kern 5pt}
&F_{\alpha\beta} = \delta_{\alpha\beta} \, F_\alpha ~, ~~~
F_\alpha =  \frac{m_{\ell_\alpha}^2}{(|c_{\ell_\alpha}| - \onehalf) \,  m_{B_\alpha}^2} ~.
\label{GHseesaw1}
\end{align}
Neutrino masses  $\{ m_{\nu_j} \}$ are eigenvalues of the matrix $FM$.
With the use of eigenvectors $\vec w_j$ the Majorana mass matrix $M$ can be expressed as (\ref{MajoranaMass1}), 
or as
\begin{align}
&M 
= F^{-1} \bigg\{ \sum_{j=1}^3 \frac{m_{\nu_j}}{n_j} \, \vec w_j  \otimes \vec w_j {}^\dagger \bigg\} F^{-1} ~.
\label{MajoranaMass3}
\end{align}
The relation (\ref{GHseesaw1}) represents a gauge-Higgs seesaw mechanism for general $M$. %Majorana mass matrix.  

The PMNS matrix $U_\PMNS$ is found to be 
\begin{align}
&U_\PMNS = \begin{pmatrix} && \cr  \vec p_1 & \vec p_2 & \vec p_3 \cr && \end{pmatrix}, \cr
\noalign{\kern 5pt}
&\vec p_j = \frac{1}{\sqrt{n_j}} \, F^{-1/2} \, \vec w_j ~, ~~~ \vec p_j {}^{\dagger} \vec p_k = \delta_{jk} ~. 
\label{PMNS3}
\end{align}
In terms of $\vec p_j$,  (\ref{GHseesaw1}) and (\ref{MajoranaMass3}) can be written  as
\begin{align}
&F^{1/2} M F^{1/2} \, \vec p_j = m_{\nu_j} \, \vec p_j ~, \cr
\noalign{\kern 5pt}
&M = F^{-1/2} \bigg\{ \sum_{j=1}^3 m_{\nu_j}\, \vec p_j  \otimes \vec p_j {}^\dagger \bigg\} F^{-1/2} ~.
\label{GHseesaw2}
\end{align}
%As $M^\dagger = M = M^t = M^*$, $\vec p_j$ becomes real.  Consequently 
The  GHU scenario predicts no CP violation phase in $U_\PMNS$.  
Neutrino oscillations with $U_\PMNS$ in the normal ordering 
with $\delta_{CP} = \pi$  are reproduced in accordance with the NuFit-6.0 analysis.  

To have a nontrivial CP violation phase in $U_\PMNS$, one needs to consider more general brane interactions and
brane masses $m^B_{\alpha\beta}$ in (\ref{branemass1}).  
In the present paper we have considered diagonal brane masses $m^B_{\alpha\beta} = \delta_{\alpha\beta} m_{B_\alpha}$.
When  $m^B_{\alpha\beta}$ is not diagonal, a general CP violation phase $\delta_{\rm CP}$ can arise but
 it becomes very difficult to solve the equations of motion explicitly.
In this regards we would like to recall that the CKM matrix in the quark sector is naturally reproduced in the 
GUT-inspired GHU with brane interactions which are diagonal in the generation space\cite{CKM2025}.
From the viewpoint of grand unification it is quite natural to imagine that brane interactions in the lepton sector
are diagonal as well.
It would be interesting to produce KK excited modes of neutrinos at the KK mass scale $m_\KK$.
Effects of lepton number violation could be seen at much lower energy scale than commonly expected as the scale of
Majorana masses can be relatively small.  Further investigation is necessary.

\section*{Acknowledgment}

The author would like to thank Yuta Orikasa for valuable comments.

% \vskip 3.cm
 
 \appendix

\section{Basis functions} 

Basis functions used for wave functions of gauge and fermion fields are given in this appendix.
For gauge fields we introduce
\begin{align}
 F_{\alpha, \beta}(u, v) &\equiv J_\alpha(u) Y_\beta(v) - Y_\alpha(u) J_\beta(v) ~, \cr
\noalign{\kern 5pt}
 C(z; \lambda) &= \frac{\pi}{2} \lambda z z_L F_{1,0}(\lambda z, \lambda z_L) ~,  \cr
 S(z; \lambda) &= -\frac{\pi}{2} \lambda  z F_{1,1}(\lambda z, \lambda z_L) ~, \cr
 C^\prime (z; \lambda) &= \frac{\pi}{2} \lambda^2 z z_L F_{0,0}(\lambda z, \lambda z_L) ~,  \cr
S^\prime (z; \lambda) &= -\frac{\pi}{2} \lambda^2 z  F_{0,1}(\lambda z, \lambda z_L)~,  \cr
\check S(z; \lambda) &= \frac{C(1; \lambda)}{S(1; \lambda)} \,  S(z; \lambda) ~, 
%\cr \noalign{\kern 5pt}
%\hat  S(z; \lambda) & =  \frac{C(1; \lambda)}{S(1; \lambda)} \,  S(z; \lambda) ~,
\label{functionA1}
\end{align}
where $J_\alpha (u)$ and $Y_\alpha (u)$ are Bessel functions of  the first and second kind.
$C(z;\lambda)$ and $S(z; \lambda)$  satisfy
\begin{align}
&- z \frac{d}{dz} \frac{1}{z} \frac{d}{dz} \begin{pmatrix} C \cr S \end{pmatrix} 
= \lambda^{2} \begin{pmatrix} C \cr S \end{pmatrix} ~,  
\label{relationA1}
\end{align}
with the boundary conditions $C = z_L$, $C'  =S = 0 $, and $S' = \lambda$ at $z=z_L$, 
and a relation $CS' - S C' = \lambda z$ holds.
%$C(z_{L} ; \lambda, z_L)  = z_{L}$, $C' (z_{L} ; \lambda, z_L)  =S(z_{L} ; \lambda, z_L)  = 0 $, 
%$S' (z_{L} ; \lambda, z_L)  = \lambda$, and $CS' - S C' = \lambda z$.
%We also make use of $\hat  S(z; \lambda) =  C(1; \lambda) S(z; \lambda) /S(1; \lambda)$.

For fermion fields with a bulk mass parameter $c$, we define 
\begin{align}
\begin{pmatrix} C_L \cr S_L \end{pmatrix} (z; \lambda,c)
&= \pm \frac{\pi}{2} \lambda \sqrt{z z_L} F_{c+\frac12, c\mp\frac12}(\lambda z, \lambda z_L) ~, \cr
\begin{pmatrix} C_R \cr S_R \end{pmatrix} (z; \lambda,c)
&= \mp \frac{\pi}{2} \lambda \sqrt{z z_L} F_{c- \frac12, c\pm\frac12}(\lambda z, \lambda z_L) ~, \cr
\begin{pmatrix} \check S_L \cr \check C_R \end{pmatrix} (z; \lambda,c)
&=\frac{C_L (1; \lambda, c)}{S_L (1; \lambda, c)} \begin{pmatrix} S_L \cr C_R \end{pmatrix} (z; \lambda,c) ~.
\label{functionA2}
\end{align}
These functions satisfy 
\begin{align}
&D_{+} (c) \begin{pmatrix} C_{L} \cr S_{L} \end{pmatrix} = \lambda  \begin{pmatrix} S_{R} \cr C_{R} \end{pmatrix}, \cr
\noalign{\kern 5pt}
&D_{-} (c) \begin{pmatrix} S_{R} \cr C_{R} \end{pmatrix} = \lambda  \begin{pmatrix} C_{L} \cr S_{L} \end{pmatrix}, \cr
\noalign{\kern 5pt}
&D_{\pm} (c) = \pm \frac{d}{dz} + \frac{c}{z} ~, 
\label{relationA2}
\end{align}
with the boundary conditions $C_{R/L} =1$, $D_-(c) C_R = D_+(c) C_L =S_{R/L} = 0$ at $z=z_{L} $, and 
$C_L C_R - S_L S_R=1$.

\vskip 1.cm

% \newpage
 
 %%%%%%%. references  %%%%%%%%%%%

% A useful Journal macro
%\def\jnl#1#2#3#4{{#1}{\bf #2} (#4) #3}
\def\jnl#1#2#3#4{{#1}{\bf #2},  #3 (#4)}

\def\Zphys{{\em Z.\ Phys.} }
\def\jssc{{\em J.\ Solid State Chem.\ }}
\def\jpsJ{{\em J.\ Phys.\ Soc.\ Japan }}
\def\ptps{{\em Prog.\ Theoret.\ Phys.\ Suppl.\ }}
\def\PTP{{\em Prog.\ Theoret.\ Phys.\  }}
\def\PTEP{{\em Prog.\ Theoret.\ Exp.\  Phys.\  }}
\def\JMP{{\em J. Math.\ Phys.} }
\def\NPB{{\em Nucl.\ Phys.} B}
\def\NP{{\em Nucl.\ Phys.} }
\def\PLB{{\it Phys.\ Lett.} B}
\def\PL{{\em Phys.\ Lett.} }
\def\PRL{\em Phys.\ Rev.\ Lett. }
\def\PRB{{\em Phys.\ Rev.} B}
\def\PRD{{\em Phys.\ Rev.} D}
\def\PRe{{\em Phys.\ Rep.} }
\def\AP{{\em Ann.\ Phys.\ (N.Y.)} }
\def\RMP{{\em Rev.\ Mod.\ Phys.} }
\def\ZPC{{\em Z.\ Phys.} C}
\def\SCI{{\em Science} }
\def\CMP{\em Comm.\ Math.\ Phys. }
\def\MPLA{{\em Mod.\ Phys.\ Lett.} A}
\def\IJMPA{{\em Int.\ J.\ Mod.\ Phys.} A}
\def\IJMPB{{\em Int.\ J.\ Mod.\ Phys.} B}
\def\EPJC{{\em Eur.\ Phys.\ J.} C}
\def\EPJP{{\em Eur.\ Phys.\ J.} Plus}
\def\PR{{\em Phys.\ Rev.} }
\def\JHEP{{\em JHEP} }
\def\JCAP{{\em JCAP} }
\def\cmp{{\em Com.\ Math.\ Phys.}}
\def\JPA{{\em J.\  Phys.} A}
\def\JPG{{\em J.\  Phys.} G}
\def\NJP{{\em New.\ J.\  Phys.} }
\def\CQG{\em Class.\ Quant.\ Grav. }
\def\ATMP{{\em Adv.\ Theoret.\ Math.\ Phys.} }
\def\ibid{{\em ibid.} }
\def\ChP{{\em Chin.Phys.}C}
\def\NCA{{\it Nuovo Cim.} A}

\renewenvironment{thebibliography}[1]
         {\begin{list}{[$\,$\arabic{enumi}$\,$]}  % {\arabic{enumi}.}
         {\usecounter{enumi}\setlength{\parsep}{0pt}
          \setlength{\itemsep}{0pt}  \renewcommand{\baselinestretch}{1.2}
          \settowidth
         {\labelwidth}{#1 ~ ~}\sloppy}}{\end{list}}

%\newpage
 \vskip 1.cm

\leftline{\Large \bf References}

%%%%%%%%%%%%% BIBLIOGRAPHY  %%%%%%%%%%%%%%%%%%%%


\begin{thebibliography}{99}
%%%%%%%%%%%%%%%%%%%%%%%%%%%%%%%%%%%%%%%%%%%%%%%

%%%%%%%%
\bibitem{Hosotani1983}
Y.\ Hosotani, 
{\it ``Dynamical mass generation by compact extra dimensions''}, 
\jnl{\PLB}{126}{309}{1983}.


\bibitem{Davies1988}
A.~T.~Davies and A.~McLachlan,
{\it ``Gauge group breaking by Wilson loops''},
\jnl{\PLB}{200}{305}{1988}.

\bibitem{Hosotani1989}
Y.\ Hosotani,  
{\it ``Dynamics of nonintegrable phases and gauge symmetry breaking''}, 
\jnl{\AP}{190}{233}{1989}.

\bibitem{Davies1989}
A.~T.~Davies and A.~McLachlan,
{\it ``Congruency class effects in the Hosotani model''},
\jnl{\NPB}{317}{237}{1989}.

\bibitem{HetrickHo1989}
J.E.~Hetrick and C-L.~Ho,
{\it ``Dynamical symmetry breaking from toroidal compactification''},
\jnl{\PRD}{40}{4085}{1989}.

\bibitem{McLachlan1990}
A.~McLachlan,
{\it ``Flux-breaking in space-times with toroidal compactification''},
\jnl{\NPB}{338}{188}{1990}.

\bibitem{Hatanaka1998}
H.\ Hatanaka, T.\ Inami, and C.S.\ Lim,
{\it ``The gauge hierarchy problem and higher dimensional gauge theories''}, 
\jnl{\MPLA}{13}{2601}{1998}.

\bibitem{Hatanaka1999}
H.\ Hatanaka,
{\it ``Matter representations and gauge symmetry breaking via compactified space''}, 
\jnl{\PTP}{102}{407}{1999}.

\bibitem{Antoniadis2001}
I.~Antoniadis, K.~Benakli and M.~Quiros,
{\it ``Finite Higgs mass without supersymmetry''},
\jnl{\NJP}{3}{20}{2001}.
% \emph{New J.~Phys.} \textbf{3}, {20} (2001).

\bibitem{Takenaga2002}
K.~Takenaga,
{\it ``Gauge symmetry breaking through the Hosotani mechanism in softly broken supersymmetric QCD''},
\jnl{\PRD}{66}{085009}{2002}.
% \emph{Phys.~Rev.}  \textbf{D66}, {085009} (2002).


\bibitem{Kubo2002}
M.\ Kubo, C.S.\ Lim and H.\ Yamashita, 
{\it ``The Hosotani mechanism in bulk gauge theories with an orbifold extra space $S^1/Z_2$''}, 
\jnl{\MPLA}{17}{2249}{2002}.

%%%%%%%

%%  SU(3) or SO(5) x U(1) gauge-Higgs %%%

%%%%%%%%%%%

\bibitem{BurdmanNomura2003}
G.~Burdman and Y.~Nomura,
{\it ``Unification of Higgs and gauge fields in five dimensions''},
\jnl{\NPB}{656}{3}{2003}.
% \emph{Nucl.~Phys.} \textbf{B656}, {3} (2003).

\bibitem{Csaki2003}
C.~Csaki, C.~Grojean and H.~Murayama,
{\it ``Standard model Higgs from higher dimensional gauge fields''},
\jnl{\PRD}{67}{085012}{2003}.
% \emph{Phys.~Rev.} \textbf{D67}, {085012} (2003).


\bibitem{Scrucca2003}
C.A.~Scrucca, M.~Serone,  and L.~Silvestrini,
{\it ``Electroweak symmetry breaking and fermion masses from extra dimensions''}, 
\jnl{\NPB}{669}{128}{2003}.

\bibitem{ACP2005}
K.~Agashe, R.~Contino and A.~Pomarol,
{\it ``The minimal composite Higgs model''}, 
\jnl{\NPB}{719}{165}{2005}.
%Nucl.\ Phys.\ {\bf B719}, 165 (2005).

\bibitem{Cacciapaglia2006}
G.~Cacciapaglia, C.~Csaki, and S.C.~Park,
{\it ``Fully radiative electroweak symmetry breaking''}, 
\jnl{\JHEP}{0603}{099}{2006}.

\bibitem{Medina2007}
A.~D.~Medina, N.~R.~Shah and C.~E.~M.~Wagner, 
{\it ``Gauge-Higgs unification and radiative electroweak symmetry breaking in warped 
extra dimensions''}, 
\jnl{\PRD}{76}{095010}{2007}.

\bibitem{HOOS2008} 
Y.~Hosotani, K.~Oda, T.~Ohnuma and Y.~Sakamura,
{\it ``Dynamical electroweak symmetry breaking in $SO(5) \times U(1)$ gauge-Higgs 
unification with top and bottom quarks''}, 
\jnl{\PRD}{78}{096002}{2008}; 
\jnl{~}{79}{079902(E)}{2009}.
%Phys.\ Rev.\  {\bf D78}, 096002 (2008); 
%{\it Erratum ibid.\ }{\bf D79}, 079902 (2009).
%{\it Erratum}-\jnl{\ibid}{{\rm D}79}{079902}{2009}.


\bibitem{Serone2010}
M.~Serone,
{\it ``Holographic methods and gauge-Higgs unification in flat extra dimensions''},
\jnl{\NJP}{12}{075013}{2010}.
% \emph{New J.~Phys.} \textbf{12}, {075013} (2010).


\bibitem{FHHOS2013} 
 S.~Funatsu, H.~Hatanaka, Y.~Hosotani, Y.~Orikasa and T.~Shimotani,
{\it ``Novel universality and Higgs decay $H\to \gamma\gamma, gg$ in the 
$SO(5) \times U(1)$ gauge-Higgs unification''}, 
\jnl{\PLB}{722}{94}{2013}.


\bibitem{Yoon2018b}
J.~Yoon and M.~E. Peskin, 
{\it ``Dissection of an $SO(5) \times U(1)$ gauge-Higgs  unification model''},
\jnl{\PRD}{100}{015001}{2019}. 


\bibitem{GUTinspired2019a}
 S.~Funatsu, H.~Hatanaka, Y.~Hosotani, Y.~Orikasa and N.~Yamatsu,
{\it ``GUT inspired $SO(5)\times U(1) \times SU(3)$ gauge-Higgs unification''}, 
\jnl{\PRD}{99}{095010}{2019}.   


\bibitem{FCNC2020a}
 S.~Funatsu, H.~Hatanaka, Y.~Hosotani, Y.~Orikasa and N.~Yamatsu,
{\it ``CKM matrix and FCNC suppression in $SO(5)\times U(1) \times SU(3)$ gauge-Higgs unification''}, 
\jnl{\PRD}{101}{055016}{2020}.   

\bibitem{GUTinspired2020b}
 S.~Funatsu, H.~Hatanaka, Y.~Hosotani, Y.~Orikasa and N.~Yamatsu,
{\it ``Effective potential and universality in GUT-inspired gauge-Higgs unification''}, 
\jnl{\PRD}{102}{015005}{2020}.   

%%%%%%%%%


%%%%%%. SO11 GHU %%%%%

\bibitem{SO11GHGU}
Y.~Hosotani and N.~Yamatsu, 
{\it ``Gauge-Higgs grand unification''}, 
\jnl{\PTEP}{2015}{111B01}{2015}, (arXiv:1504.03817 [hep-ph]); 

\bibitem{Furui2016}
A.~Furui, Y.~Hosotani, and N.~Yamatsu, 
{\it ``Toward realistic gauge-Higgs grand  unification''}, 
\jnl{\PTEP}{2016}{093B01}{2016}, (arXiv:1606.07222 [hep-ph]).

\bibitem{HosotaniYamatsu2017}
Y.~Hosotani and N.~Yamatsu, 
{\it ``Gauge-Higgs seesaw mechanism in 6-dimensional  grand unification''}, 
\jnl{\PTEP}{2017}{091B01}{2017}.
%\emph{Prog.~Theoret.~Exp.~Phys.}  \textbf{2017}, 091B01 (2017).

\bibitem{HosotaniYamatsu2018}
Y.~Hosotani and N.~Yamatsu, 
{\it ``Electroweak symmetry breaking and mass spectra  in six-dimensional  gauge-Higgs grand unification''}, 
\jnl{\PTEP}{2018}{023B05}{2018}.


%%%%%%%%%%


%%%%%%
\bibitem{Funatsu2017a}
S.~Funatsu, H.~Hatanaka, Y.~Hosotani, and Y.~Orikasa, 
{\it ``Distinct signals of  the gauge-Higgs unification in $e^+e^-$ collider experiments''}, 
\jnl{\PLB}{775}{297}{2017}.   

\bibitem{Yoon2018a}
J.~Yoon and M.~E. Peskin, 
{\it ``Fermion pair production in $SO(5) \times U(1)$  gauge-Higgs unification models''}, 
arXiv:1811.07877.

\bibitem{Funatsu2019a}
S.~Funatsu, 
{\it ``Forward-backward asymmetry in the gauge-Higgs unification at the International Linear Collider''},
\jnl{\EPJC}{79}{854}{2019}.   


\bibitem{GUTinspired2020c}
 S.~Funatsu, H.~Hatanaka, Y.~Hosotani, Y.~Orikasa and N.~Yamatsu,
{\it ``Fermion pair production at $e^- e^+$ linear collider experiments
in GUT inspired gauge-Higgs unification''}, 
\jnl{\PRD}{102}{015029}{2020}.

\bibitem{Irles2024}
A.~Irles, J.P.~M\'{a}rquez, R.~P\"{o}schl, F.~Richard, A.~Saibel, H.~Yamamoto and N.~Yamatsu,
{\it ``Probing gauge-Higgs unification models at the ILC with quark-antiquark forward-backward asymmetry 
at center-of-mass energies above the $Z$ mass''},
\jnl{\EPJC}{84}{537}{2024}.  
% \emph{Eur.~Phys.~J.} \textbf{C84}, {537} (2024).

%%%%%%%%%%%%%%%%%%%%%%
 \bibitem{Funatsu2023a}
 S.~Funatsu, H.~Hatanaka, Y.~Orikasa, and N.~Yamatsu, 
 {\it ``Single Higgs boson production at electron-positron colliders in gauge-Higgs unification"},
 \jnl{\PRD}{107}{075030}{2023}.

\bibitem{Yamatsu2023}
N.~Yamatsu,  S.~Funatsu, H.~Hatanaka, Y.~Hosotani,  and Y.~Orikasa
{\it ``$W$  and $Z$ boson pair production at electron-positron colliders  in gauge-Higgs unification''}, 
\jnl{\PRD}{108}{115014}{2023}.
% arXiv:2309.01132 [hep-ph]. (To appear in \PRD.)

%%%%%%. W mass %%%%%%%%%%

\bibitem{Wmass2023}
 Y.~Hosotani,  S.~Funatsu, H.~Hatanaka,Y.~Orikasa and N.~Yamatsu,
{\it ``$W$ boson mass in gauge-Higgs unification''}, 
\jnl{\PRD}{108}{115036}{2023}.


\bibitem{CDF2022}
T.~Aaltonen et al.\ (CDF Collaboration),
{\it ``High-precision measurement of the $W$ boson  mass with the CDF II detector''},
\jnl{\SCI}{376}{170}{2022}.

% \bibitem{ATLAS2023}
% The ATLAS Collaboration,
% {\it ``Improved $W$ boson mass measurement using $\sqrt{s} = 7\,$TeV proton-proton collisions with the ATLAS detector''},
% ATLAS CONF Note, 
% Report No. ATLAS-CONF-2023-004.

% \bibitem{CMS2024}
% The CMS Collaboration,
% {\it ``High-precision measurement of the  $W$ boson mass  with the CMS experiment at the LHC''},
% Report No. CMS-SMP-23-002. (arXiv:2412.13872 [hep-ex])

%%%%%%%%%%%%%%%%%%%


\bibitem{YHbook}
Y.~Hosotani,
{\it ``An Introduction to Gauge-Higgs Unification''},
(World Scientific Publishing Company, Singapore, 2025), 
ISBN 978-981-98-0097-1, doi.org/10.1142/14045.

%%%%%%%%%%%%%%%%%%%%%%

%%%%%%%%%%.  anomaly flow. %%%%%%%%

\bibitem{AnomalyFlow1}
S.~Funatsu, H.~Hatanaka,  Y.~Hosotani,  Y.~Orikasa and N.~Yamatsu,
{\it ``Anomaly flow by an Aharonov-Bohm phase''},
\jnl{\PTEP}{2022}{043B04 }{2022}, (arXiv:2202.01393 [hep-ph]).
% \emph{Prog.~Theoret.~Exp.~Phys.}  \textbf{2022}, 043B04 (2022).

\bibitem{AnomalyFlow2}
Y.~Hosotani,   
{\it ``Universality in anomaly flow''}, 
\jnl{\PTEP}{2022}{073B01 }{2022}, (arXiv:2205.00154 [hep-th]).
% \emph{Prog.~Theoret.~Exp.~Phys.}  \textbf{2022}, 073B01 (2022).

\bibitem{AnomalyFlow3}
Y.~Hosotani,   
{\it ``Holography  in anomaly flow in orbifold gauge theory''}, 
\jnl{{\it PoS} CORFU }{2022}{120}{2023}, (arXiv:2309.01132 [hep-ph]).

%%%%%%%%%%%%%%%%

%%%%%%%%. grand unification.  %%%%%%%

\bibitem{LimMaru2007}
C.S.~Lim and N.~Maru,
{\it ``Towards a realistic grand gauge-Higgs unification''},
\jnl{\PLB}{653}{320}{2007}.   
% \emph{Phys.~Lett.} \textbf{B653}, {320} (2007).

\bibitem{Kojima2011}
K.~Kojima, K.~Takenaga and T.~Yamashita,
{\it ``Grand gauge-Higgs unification''},
\jnl{\PRD}{84}{051701(R)}{2011};
% \emph{Phys.~Rev.} \textbf{D84}, {051701(R)} (2011).
% \bibitem{Kojima2017}
% K.~Kojima, K.~Takenaga and T.~Yamashita,
{\it ``Gauge symmetry breaking patterns in an $SU(5)$ grand gauge-Higgs unification model''},
\jnl{\PRD}{95}{015021}{2017}.
% \emph{Phys.~Rev.} \textbf{D95}, {015021} (2017).

\bibitem{Englert2020}
C.~Englert, D.J.~Miller and D.D.~Smaranda,
{\it ``Phenomenology of GUT-inspired gauge-Higgs unification''},
\jnl{\PLB}{802}{135261}{2020};
% \emph{Phys.~Lett.} \textbf{B802}, {135261} (2020).
% \bibitem{Englert2020b}
% C.~Englert, D.J.~Miller and D.D.~Smaranda,
{\it ``The Weinberg angle and 5D RGE effects in a $SO(11)$ GUT theory''},
\jnl{\PLB}{807}{135548}{2020}.
% \emph{Phys.~Lett.} \textbf{B807}, {135548} (2020).

\bibitem{Angelescu2022}
A.~Angelescu, A.~Bally, S.~Blasi and F.~Goertz,
{\it ``Minimal $SU(6)$ gauge-Higgs grand unification''},
\jnl{\PRD}{105}{035026}{2022}.
% \emph{Phys.~Rev.} \textbf{D105}, {035026} (2022).

\bibitem{Maru2022}
N.~Maru, H.~Takahashi and Y.~Yatagai,
{\it ``Gauge coupling unification in simplified grand gauge-Higgs unification''},
\jnl{\PRD}{106}{055033}{2022}.
% \emph{Phys.~Rev.} \textbf{D106}, {055033} (2022).

\bibitem{Angelescu2023}
A.~Angelescu, A.~Bally,  F.~Goertz and S.~Weber,
{\it ``$SU(6)$ gauge-Higgs grand unification: Minimal viable models and flavor''},
\jnl{\JHEP}{04}{012}{2023}.  
% \emph{J.~High Energy Phys.} \textbf{04}, {012} (2023).

\bibitem{MaruNago2024}
N.~Maru and R.~Nago,
{\it ``New models of $SU(6)$ grand gauge-Higgs unification''},
\jnl{\JHEP}{11}{035}{2024}.  
% arXiv:2405.07463.


\bibitem{MaruNago2025}
N.~Maru and R.~Nago,
{\it ``Electroweak symmetry breaking in $Sp(6)$  gauge-Higgs unification model''},
\jnl{\NPB}{1014}{116863}{2025}.  
% arXiv:2411.02808.



%%%%%%%%%%%%%%%%%%

\bibitem{CKM2025}
 Y.~Hosotani,  S.~Funatsu, H.~Hatanaka, Y.~Orikasa and N.~Yamatsu,
{\it ``Origin of CKM matrix and natural FCNC suppression in gauge-Higgs unification''}, 
\jnl{\PRD}{111}{095023}{2025}.


%%%%%%%%%%%%%%%%

\bibitem{Pontecorvo1958}
B.~Pontecorvo,
{\it ``Inverse beta processes and nonconservation of lepton charge''},
\jnl{{\it Sov.\ Phys.\ JETP }}{7}{172}{1958}.
% {\it ``Mesonium and antimesonium''},
% \jnl{{\it Sov.\ Phys.\ JETP }}{6}{429}{1958}.

\bibitem{MNS1962}
Z.~Maki, M.~Nakagawa and S.~Sakata,
{\it ``Remarks on the unified model of elementary particles''}, 
\jnl{\PTP}{28}{870}{1962}.

% \bibitem{Cabibbo1963}
% N.~Cabibbo,
% {\it ``Unitary symmetry and leptonic decays''}, 
% \jnl{\PRL}{10}{531}{1963}.


\bibitem{KM1973}
M.~Kobayashi and T.~Maskawa,
{\it ``CP violation in the renormalizable theory of weak interaction''}, 
\jnl{\PTP}{49}{652}{1973}.


\bibitem{NuFit2024}
I.~Esteban, M.C.~Gonzalez-Garcia, M.~Maltoni, I.~Martinez-Soler, J.P.~Pinheiro, T.~Schwetz,
{\it ``NuFit-6.0: Updated global analysis of three-flavor neutrino oscillations''}, 
\jnl{\JHEP}{12}{216}{2024}.




%%%%%%%%
\bibitem{RS1}
L.\ Randall and R.\ Sundrum,
{\it ``A large mass hierarchy from a small extra dimension''}, 
\jnl{\PRL}{83}{3370}{1999}.

%%%%%%%%%%%%%%%


%%%%%%% twisted gauge  %%%%%%

\bibitem{Falkowski2007}
A.~Falkowski,
{\it ``Holographic pseudo-Goldstone boson''},
\jnl{\PRD}{75}{025017}{2007}.

\bibitem{HS2007}
Y.\ Hosotani and Y.\ Sakamura,
{\it ``Anomalous Higgs couplings in the $SO(5) \times U(1)_{B-L}$ gauge-Higgs
unification in warped spacetime''},
\jnl{\PTP}{118}{935}{2007}.

%%%%%%%%. Seesaw mechanism.  %%%%%%%%%

\bibitem{HasegawaLim2018}
K.~Hasegawa and C.S.~Lim,
{\it ``Majorana neutrino masses in the scenario of gauge-Higgs unification''},
\jnl{\PTEP}{2018}{073B01 }{2018}.


\bibitem{Minkowski1977}
P.\ Minkowski,
{\it ``$\mu \go e \gamma$ at a rate of one out of $10^9$ muon decays?''}, 
% \emph{Phys.~Lett.} \textbf{B67}, {421} (1977).
 \jnl{\PLB}{67}{421}{1977}.


\bibitem{Yanagida1979}
T.\ Yanagida, 
{\it ``Horizontal gauge symmetry and masses of neutrinos''}, 
\jnl{\it Conf.\ Proc.\ C~}{7902131}{95}{1979}.
%. https://inspirehep.net/literature/143150
% in {\it Proceedings of Workshop on Unified Theory and Baryon Number of the Universe},
% edited by O.\ Sawada and A.\ Sugamoto (KEK, Japan, 1979).


\bibitem{GellMann1980}
M.\ Gell-Mann, P.\ Ramond and R.\ Slansky, 
{\it ``Complex spinors and unified theories''}, 
in {\it Supergravity}, 
edited by P.\ van Nieuwenhuizen and D.\ Freedman (North-Holland, Amsterdam, 1980) p.\ 317.


\bibitem{Mohapatra1986}
R.N.\ Mohapatra, 
{\it ``Mechanism for understanding small neutrino mass in superstring theories''}, 
% \emph{Phys.~Rev.~Lett.} \textbf{56}, {561} (1986).
\jnl{\PRL}{56}{561}{1986}.


\bibitem{MohapatraValle1986}
R.N.\ Mohapatra and J.W.F.\ Valle, 
{\it ``Neutrino mass and baryon-number nonconservation in superstring models''},  
%\emph{Phys.~Rev.} \textbf{D34}, {1642} (1986).
\jnl{\PRD}{34}{1642}{1986}.


\bibitem{Shafi2004}
S.J.\ Huber and Q.\ Shafi, 
{\it ``Seesaw mechanism in warped geometry''},  
\jnl{\PLB}{583}{293}{2004}.

\bibitem{Agashe2016}
K.\ Agashe, S.\ Hong and L.\ Vecchi,
{\it ``Warped seesaw mechanism is physically inverted''},  
\jnl{\PRD}{94}{013001}{2016}.


%%%%%%%%%%%%%%%%%%%%%%%


\end{thebibliography}
\end{document}